\documentclass[letterpaper,10pt]{article}
\usepackage{jheppub} % for details on the use of the package, please
\usepackage{bm} 
\usepackage{slashed} 
\usepackage{amsfonts} 
\usepackage{amsmath} 
\usepackage{bbm} 
\usepackage{amssymb} 
\usepackage{graphicx} 
\usepackage[usenames,dvipsnames]{xcolor} 
\newcommand{\aeq}{\begin{equation}} 
\newcommand{\ceq}{\end{equation}} 
\newcommand{\aec}{\begin{eqnarray}} 
\newcommand{\cec}{\end{eqnarray}} 
\newcommand{\ase}{\begin{subequations}} 
\newcommand{\cse}{\end{subequations}} 

\renewcommand{\(}{\left(} 
\renewcommand{\)}{\right)}

\renewcommand{\a}{\alpha} 
\renewcommand{\b}{\beta} 
\renewcommand{\c}{\chi} 
\newcommand{\m}{\mu} 
\newcommand{\n}{\nu} 
\renewcommand{\o}{\omega} 
\newcommand{\g}{\gamma} 
\renewcommand{\d}{\delta} 
\newcommand{\h}{\eta} 
\newcommand{\z}{\zeta}

\newcommand{\y}{\psi} 
\renewcommand{\l}{\lambda} 
\newcommand{\s}{\sigma} 
\renewcommand{\r}{\rho} 
\renewcommand{\k}{\kappa} 
\newcommand{\q}{\theta} 
\newcommand{\e}{\epsilon} 
 
\renewcommand{\t}{\tau} 
\newcommand{\p}{\pi} 

\newcommand{\G}{\Gamma} 
\renewcommand{\P}{\Pi}

\newcommand{\D}{\Delta}

\newcommand{\pd}{\partial} 

\title{\boldmath Second order theory of $(j,0)\oplus (0,j)$ single high spins as Lorentz tensors}

\author{E. G. Delgado-Acosta}
\author{and M. Kirchbach}

\affiliation{Instituto de F\'{\i}sica, Universidad Aut\'onoma de San Luis Potos\'{\i},\\ 
Av. Manuel Nava 6, San Luis Potos\'{\i}, S.L.P. 78290, M\'exico}

\emailAdd{german@ifisica.uaslp.mx}
\emailAdd{mariana@ifisica.uaslp.mx}

\abstract{We show that higher order differential equations and matrix spinor calculus are completely 
avoidable in the description of pure high spin-$j$ Weinberg-Joos states, $(j,0)\oplus (0,j)$. 
The case is made on the example of  $\left(\frac{3}{2},0\right)\oplus \left(0,\frac{3}{2}\right)$, 
for the sake of concreteness and without loss of generality. Namely, we use as a vehicle for the 
aforementioned covariant single spin-$\frac{3}{2}$ description the direct sum of 
$\left(\frac{3}{2},0\right)\oplus \left(0,\frac{3}{2}\right)$ with the Dirac field, 
$\psi\simeq \left(\frac{1}{2},0\right)\oplus \left(0,\frac{1}{2}\right)$, on the one side, and 
$\left(\frac{1}{2},1\right)\oplus \left(1,\frac{1}{2}\right)$, on the other, 
which amounts to the antisymmetric tensor of second rank with Dirac spinor components, 
$\Psi_{\left[\mu\nu\right]}=B_{\left[\mu\nu\right]}\otimes\psi$. 
The $\left(\frac{3}{2},0\right)\oplus \left(0,\frac{3}{2}\right)$ sector of interest is then tracked 
down in two steps. First we search for spin- $\frac{3}{2}$ by means of a covariant spin projector 
constructed from the Casimir invariants of the Poincar\'e algebra, the squared four momentum, $P^2$, 
and the squared Pauli-Lubanski vector, ${\mathcal W}^2$. This projector is second order in the momenta. 
Afterwords we identify the wanted irreducible representation space  by means of a momentum independent 
(static) projector designed on the basis of the Casimir invariants of the Lorentz algebra. The latter 
projectors have  the property to unambiguously identify any irreducible $so(1,3)$ subspace of any Lorentz 
tensor and without rising the order of the differential equation. In this fashion, a  Lagrangian that is 
second order in the momenta is furnished. The method proposed correctly reproduces the  electromagnetic 
multipole moments earlier calculated for $\left( \frac{3}{2},0\right)\oplus \left(0, \frac{3}{2}\right)$ 
in treating it in the standard way as eight dimensional spinor. We furthermore calculate Compton scattering 
off the pure spin-$\frac{3}{2}$ under discussion, and show that the differential cross section satisfies 
unitarity in forward direction for a gyromagnetic ratio of $g=\frac{2}{3}$. This result hints on the possible 
validity of Belinfante's conjecture for pure spin-states, while the natural value of $g=2$ seems more likely 
to characterize the highest spins in the  Rarita-Schwinger representation spaces. The scheme straightforwardly 
extends to any $(j,0)\oplus (0,j)$ Weinberg-Joos state and brings the advantage of avoiding rectangular matrix 
couplings between states of different spins, replacing them by simple Lorentz contractions.}


\begin{document} 
\maketitle
\flushbottom

%%%%%%%%%%%%%%%%%%%%%%%%%%%%%%%%%%%%%%%%% 
\section{Methods for high-spin descriptions - introductory remarks}\label{sec1} 
%%%%%%%%%%%%%%%%%%%%%%%%%%%%%%%%%%%%%%%%%%%%%%%%%%%%%%%%%%%%%%%%%%%%%%%%%%%%%% 

%%%%%%%%%%%%%%%%%%%%%%%%%%%%%%%%%%%%%%%%% 

Particles of high-spins  \cite{Weinberg:1995mt} continue being among the most 
enigmatic  challenges in contemporary 
theoretical physics. The difficulties in their descriptions, both at the 
classical-, and the quantum-field theoretical 
levels, are well known and take their origin from the circumstance  that such 
particles are most naturally described 
by  differential equations of orders  twice their  respective  spins 
\cite{Weinberg64},\cite{Joos}. Higher-order theories are  difficult to tackle 
and various strategies have been elaborated over the years to lower the order 
of the corresponding differential equations, the linear 
ones by Rarita-Schwinger \cite{RS} being the most popular so far. However, the 
latter framework is plagued by various 
inconsistencies, the acausal propagation within an electromagnetic 
environment \cite{Velo:1970ur}, the violation of unitarity in 
Compton scattering in the ultraviolet in schemes with minimal gauge couplings 
\cite{Ferrara}, 
and the violation of Lorentz-symmetry  upon quantization, being the most 
serious ones. 
In parallel, also second order spin-$\frac{1}{2}$ \cite{Hostler}, 
\cite{Morgan}, \cite{Vaquero}, \cite{Rene} and spin-$\frac{3}{2}$ 
\cite{Napsuciale:2006wr} 
fermion theories have been developed  by different authors and shown to 
provide a reasonable compromise between the rigorous linear-- 
and the  natural higher-order descriptions in so far as they were able to 
circumvent the acausality problem 
simultaneously with the violation of unitarity in Compton scattering 
\cite{DelgadoAcosta:2009ic}. However, 
for spins higher than $\frac{3}{2}$ no second order theory has been developed 
so far. It is the goal of the present work to fill this gap. 
The interest in such a study is motivated by the observation that particles 
with spin-$j$ transforming 
according to distinct representation spaces  of the Lorentz algebra describe 
particles of different physical properties. 
{}For example, because of the representation dependence of the boost operator, 
the  electromagnetic quadrupole and octupole moments of fundamental particles 
with spin-$\frac{3}{2}$ transforming 
in the four-vector spinor come out  different from those of particles 
transforming  as 
$\left( \frac{3}{2},0\right) \oplus \left(0,\frac{3}{2} \right)$ 
\cite{DelgadoAcosta:2012yc}. 
Same holds valid regarding spin-$1$ in the four-vector, $\left( \frac{1} 
{2},\frac{1}{2}\right)$, 
versus the anti-symmetric tensor, $\left( 1,0\right)\oplus\left( 0,1\right)$. 
In view of the expected production of new particles in the experiments run by 
the Large Hadron Collider 
it is important to have at ones disposal a reliable and comfortable to deal 
with universal calculation scheme 
of any high spin,  be it bosonic, or fermionic.  The present study is devoted 
to the elaboration of such a scheme. 
In the current section we present a concise 
review of the persisting techniques in high-spin description of frequent use 
in low and intermediate energy physics 
emphasizing on their differences and similarities. In due course we  suggest 
our announced  strategy for any high-spin second order formalism. 
It is based upon properly constructed Lorentz tensors for bosons, or Lorentz 
tensor-spinors for fermions, and the employment of 
momentum independent projectors on irreducible spaces (``irreps'') of the 
Lorentz algebra 
$so(1,3)$ in combination with second order mass-$m$ and spin-$j$ projectors 
constructed from the two invariants of the Poincar\'e algebra, the squared 
four momentum $P^2$ and the squared Pauli-Lubanski vector, 
${\mathcal W}^2$.\\ 

\underline{\it Totally symmetric tensor- and tensor-spinor representation 
spaces:}\\ 

\noindent 
Particles of high-spins, $j>\frac{1}{2}$, have been so far most frequently 
described in the 
literature in terms of multi-spin-parity $so(1,3)$ 
representation spaces 
given by totally symmetric Lorentz tensors of the type, 
\begin{equation} 
A_{\mu_1...\mu_j}\simeq \left(\frac{j}{2},\frac{j}{2}\right), 
\end{equation} 
for bosons, or Lorentz tensor-spinors, 
\begin{equation} 
\psi_{\mu_1...\mu_{j-\frac{1}{2}}}\simeq \left(\frac{j-\frac{1}{2}}{2}, 
\frac{j-\frac{1}{2}}{2}\right)\otimes \left[ 
\left(\frac{1}{2},0\right)\oplus \left(0,\frac{1}{2}\right)\right], 
\end{equation} 
for fermions, respectively \cite{RS}, \cite{Weinberg:1995mt}. 
They  have been associated with 
the highest spins in the spaces under discussion, while their lower spin 
companions have been treated as redundant and had to be projected out in order 
to ensure the correct number 
of physical degrees of freedom for 
spin-$j$ description.\\ 

\underline{\it Rarita-Schwinger's wave equations:}\\ 

\noindent 
The wave equations for the latter case are obtained from the requirement that 
in any Lorentz index the 
field satisfies the Dirac equation, supplemented by certain auxiliary 
conditions according to 
\begin{eqnarray} 
(i\slashed{\partial}-m) 
\psi_{\mu_1...\mu_i...\mu_{j-\frac{1}{2}}}&=&0,\nonumber\\ 
\gamma^{\mu_i}\psi_{\mu_1...\mu_i...\mu_{j-\frac{1}{2}}}=0, &\quad& 
\partial^{\mu_i}\psi_{\mu_1...\mu_i...\mu_{j-\frac{1}{2}}}=0. 
\end{eqnarray} 
Specifically for spin-$\frac{3}{2}$ one has to consider the four-vector-- 
spinor, $\psi_{\mu}$, 
\begin{equation} 
 \psi_{\mu}=A_{\mu}\otimes\psi\simeq\left(  \frac{1}{2},\frac{1}{2}\right) 
\otimes\left[  \left(  \frac{1}{2},0\right)  \oplus\left(  \frac{1}% 
{2},0\right)  \right]\,, 
\label{RS_tensors} 
\end{equation} 
the direct product between the four vector, $A_{\mu}$, and the Dirac spinor, 
$\psi$, and solve the system of three linear differential equations 
\begin{eqnarray} 
(i\slashed\partial  -m)\psi_{\mu}  &  =&0\,,\label{RS_Dirac}\\ 
\gamma^{\mu}\psi_{\mu}  &  =&0\,,\label{RS_Proca}\\ 
\partial^{\mu}\psi_{\mu}  &  =& 0\, . \label{RS_What}% 
\label{RS-eqs} 
\end{eqnarray} 

\underline{\it Multi-component representation spaces:}\\ 

\noindent 
Alternatively, spin-$j$ particles can be described in terms of the single spin 
valued non-tensorial representation spaces, 
\begin{eqnarray} 
\psi^{(j)}_{B} &\simeq & (j,0)\oplus (0,j),\quad 
B \in \left[1, 2(2j+1) \right]. 
\label{WJ-spinors} 
\end{eqnarray} 

\underline{\it  Weinberg-Joos wave equations:}\\ 

\noindent 
The wave functions of particles described in terms of the representation 
spaces in (\ref{WJ-spinors}) satisfy higher order differential  equations according to, 

\begin{eqnarray} 
\left(i^{2j}\left[ 
\gamma_{\mu_1\mu_2...\mu_{2j}}\right]_{AB}\partial^{\mu_1}\partial^{\mu_2}... 
\partial^{\mu_{2j}}-m^{2j}\delta_{AB}\right)\psi^{(j)}_B(x)&=&0, 
\label{WJ-eqs} 
\end{eqnarray} 
where $\psi_B^{(j)}(x)$ is the $2(2j+1)$-component field 
$(j,0)\oplus (0,j)$, $\left[ \gamma_{\mu_1\mu_2...-\mu_{2j}}\right]_{AB}$ are 
the elements of the 
generalized Dirac Hermitean matrices of dimensionality 
$\left[2(2j+1)\right]\times \left[2(2j+1)\right]$, which 
transform as Lorentz tensors of rank-$2j$. 
The complete sets of such matrices have been extensively studied in the 
literature 
for the purpose of constructing all the possible field bilinears needed in the 
definitions of the 
generalized  currents, both  transitional  and diagonal \cite{Sankar}-- 
\cite{Eeg3}. 
Though well elaborated, this so called Weinberg-Joos formalism has attracted 
comparatively less 
attention 
than the linear Rarita-Schwinger framework, mainly because of the difficult to 
handle higher order of the 
differential equations involved, on the one side, 
and the high dimensionality of the generalized Dirac matrices, on the other. 

A possibility to put the Weinberg-Joos formalism on comparable  footing with 
that by  Rarita-Schwinger 
is to find a way to describe  single-spin representation spaces  by means of 
Lorentz-tensors. 
So far no such method for any spin, both integer and fractional, has been 
suggested in the literature, a 
circumstance that has presented over the years a serious obstacle in the 
employment of $(j,0)\oplus (0,j)$-fields 
in the description  of physical processes. Instead, the Rarita-Schwinger 
spaces have been given the clear preference 
precisely for admitting comfortable couplings to the fundamental photon-proton 
system. 
However, particles of equal spins residing in different 
representation spaces can throughout be characterized by distinct physical 
properties such as electromagnetic multipole moments, 
Compton scattering cross sections etc., this because of the non-trivial 
differences in the structures of the respective  boost operators 
\cite{DelgadoAcosta:2012yc},\cite{DelgadoAcosta:2013}. 
To study such differences it is important to have at ones disposal a tool 
for the efficient description of single-spins in terms of 
Lorentz tensors. It is the goal of the present study to provide such a tool. 
\\ 

\underline{\it Pairwise anti-symmetric tensor-spinor  representation 
spaces:}\\ 

\noindent 

In \cite{Niederle}  a linear approach in analogy to  the 
Rarita-Schwinger formalism has been elaborated for fractional high spins $j$ 
as part of an anti-symmetric (in pairs of indexes) tensor of rank $(2j-1)$, 
\begin{equation} 
\psi^{\left[ \mu_1\mu_2\right]... \left[\mu_i\nu_i \right].... \left[ \mu_{j- 
\frac{1}{2}} \nu_{j-\frac{1}{2}}\right]}\simeq 
\left[ \left( j-\frac{1}{2},0\right)\oplus \left( 0,j-\frac{1} 
{2}\right)\right]\otimes 
\left[\left( \frac{1}{2},0\right) \oplus \left(0,\frac{1}{2} \right) \right]. 
\label{NN_1} 
\end{equation} 

\underline{\it Niederle-Nikitin's fermion wave equations:}\\ 

\noindent 
The corresponding wave equations are then 
\begin{equation} 
(\slashed{p} -m)\psi^{\left[ \mu_1\mu_2\right]... \left[\mu_i\nu_i \right].... 
\left[ \mu_{j-\frac{1}{2}} \nu_{j-\frac{1}{2}}\right]} 
-\frac{1}{4j}\Sigma_{\mathcal P}\left(\gamma^{\mu_1}\gamma^{\nu_1}- 
\gamma^{\nu_1} \gamma^{\mu_1}\right)p_\lambda \gamma_\sigma 
\psi^{\left[ \lambda \sigma \right]... \left[\mu_i\nu_i \right].... \left[ 
\mu_{j-\frac{1}{2}} \nu_{j-\frac{1}{2}}\right]}=0, 
\label{NN_2} 
\end{equation} 
where ${\mathcal P}$ denotes permutations of $\mu_i$ and $\nu_i$, as well as 
of  $\left[ \mu_k\nu_k\right]$ 
with $\left[ \mu_j\nu_j\right]$. 

We here take a different path. Namely, we embed single  spin-$j$ Weinberg-Joos 
states, $(j,0)\oplus (0,j)$, 
into direct sums of properly selected irreducible $so(1,3)$ representation 
spaces which are large enough as to 
allow to be equipped by Lorentz, and if needed, separate Dirac indexes. 
Then we identify  the state of our interest in a two step procedure. 
First we search through the aforementioned direct sum for the spin of our 
interest 
using a covariant spin projector constructed from the 
Casimir invariants of the Poincar\'e algebra, the squared four momentum, 
$P^2$, and the squared Pauli-Lubanski vector, ${\mathcal W}^2$. 
Afterwords  we search for the irreducible representation space by means of a 
momentum independent (static) 
projector designed on the basis of the Casimir invariants of the Lorentz 
algebra. 
As long as the  covariant spin  projector is second order in the momenta, the 
emerging 
Lagrangian  is of  second order too. 
In this fashion, a second order formalism for any single-spin valued Weinberg- 
Joos state is furnished. 

The paper is organized as follows. In the next section we present our 
suggested 
method. In section \ref{sec:3} we  find wave equations and Lagrangians 
for all the spins residing in the antisymmetric tensor spinor and transforming 
according to one of the irreducible representation spaces, 
to be termed to in the following by {\em irrep\/}. 
In section \ref{sec:4} we construct all the explicit degrees of freedom spanning the 
tensor-spinor space under consideration. 
In section \ref{sec:5} we gauge the spin-$\frac{1}{2}$, and spin-$\frac{3}{2}$ 
Lagragians, 
find the electromagnetic currents of interest, and 
calculate the associated electromagnetic multipole moments. We show that the 
observables calculated in this fashion 
reproduce  those earlier obtained in \cite{DelgadoAcosta:2012yc} from 
considering pure  spin-$\frac{3}{2}$ state in the standard way as an eight- 
dimensional spinor. 
Also there we show that the pure spin-$\frac{1}{2}$ sector of the 
antisymmetric tensor spinor describes a genuine Dirac particle, while 
the properties of the particles in $\left(\frac{1}{2},1 \right)\oplus 
\left(1,\frac{1}{2} \right)$ are same as those of the corresponding 
part of the Rarita-Schwinger four vector spinor. 
Section \ref{sec:6} is devoted to the evaluation of Compton scattering off  $\left( 
\frac{3}{2},0\right)\oplus \left(0,\frac{3}{2} \right)$. 
There we report on finite forward differential cross section in the ultraviolet for the 
gyromagnetic ratio taking the value of the inverse spin, 
i.e. for $g=\frac{2}{3}$ and in accord with Belinfante's conjecture. The paper 
closes with brief conclusions. 

%%%%%%%%%%%%%%%%%%%%%%%%%%%%%%%%%%%%%%%%%%%%%%%%%%%%%%%%%%%%%%%%%%%%%%%%%%%%%% 

%%%%%%%%%%%%%%%%%%%%%%%%%%%%%%%%%%%%%%%%% 
%%%%%%%%%%%%%%%%%%%%%%%%%%%%%%%%%%%%%%%%%%%%%%%%%%%%%%%%%%%%%%%%%%%%%%%%%%%%%% 

%%%%%%%%%%%%%%%%%%%%%%%%%%%%%%%%%%%%%%%%% 

\section{\label{sec:2}High-spins within the relativistic  invariants (RInS) method: 
Covariant spin-irrep  projectors} 

The method for high-spin description advocated in this work is based upon 
representation spaces which are 
different from those used in the three schemes highlighted in the 
introduction. 
While the representation spaces underlying the Weinberg-Joos formalism are of 
non-tensorial nature, 
those underlying the Rarita-Schwinger framework are totally symmetric tensors. 
Finally, the 
  Niederle-Nikitin method is based on tensor-spinors that are antisymmetric in 
pairs of indexes, 
though symmetric with respect to the pairs. We here instead use Lorentz 
tensors of mixed symmetries. 
Then the irreducible tensor sector is tracked down by a projector operator 
designed on the basis of 
the momentum independent invariants of the Lorentz algebra, while the spin is 
identified by means of a 
projector constructed from the momentum dependent invariants of the Poincar\'e 
algebra. In the following, this method will be 
occasionally termed to as RInS-formalism.\\ 

\underline{\it Mixed-symmetric tensor (bosonic) and  tensor-spinor 
(fermionic) representation spaces:}\\ 

\noindent 
Our idea is to embed  $(j,0)\oplus (0,j)$ 
carrier spaces of the Lorentz algebra $so(1,3)$ 
into finite direct sums of  properly chosen auxiliary irreducible 
representation spaces with the 
aim to end up with a reducible representation space that is 
large enough as to allow  to be equipped by  Lorentz-- (and if needed, 
separate  Dirac) indexes, i.e. 
\begin{eqnarray} 
  \Psi_{\mu_1,..\mu_t}&\simeq& 
\left[(j,0)\oplus (0,j)\right] \oplus 
\Sigma _{(k,l)} \, n_{(kl)} \left[ 
\left(j_k,j_l\right)\oplus \left(j_l,j_k\right)\right]. 
\label{chudo_1} 
\end{eqnarray} 
Alternatively, spin-$j$ can be described in terms of 
two-spin valued representation spaces, $\left(j\mp \frac{1}{2},\frac{1}{2} 
\right)\oplus \left(\frac{1}{2},j\mp \frac{1}{2} \right)$, 
in which case the carrier tensor-spinor can be designed as, 
\begin{eqnarray} 
\Psi_{\mu_1,..\mu_r}&\simeq& 
\left(s,\frac{1}{2}\right) 
\oplus \left( 
\frac{1}{2}, s 
\right)\oplus \Sigma _{(k,l)} \, n_{(kl)}\left[ \left(j_k,j_l\right)\oplus 
\left(j_l,j_k\right)\right],\quad j=s\pm \frac{1}{2} , 
\label{chudo_2} 
\end{eqnarray} 
where  the generic coefficients $n_{(kl)}$ stand for  the multiplicity of the 
attached  states required to complete 
the Lorentz tensor (tensor-spinor) under construction. Contrary to the totally 
symmetric Rarita-Schwinger tensors 
in (\ref{RS_tensors}), 
the tensors in (\ref{chudo_1})-(\ref{chudo_2}) can be antisymmetric in some of 
the indexes 
  and symmetric in others.\\ 

{}For example, pure spin-$\frac{3}{2}$ can be embedded into the totally 
antisymmetric 
tensor of second rank with Dirac spinor components, 
$\Psi_{\left[ \mu\nu\right] }$, a representation space that is reducible 
according to 
\begin{eqnarray} 
\Psi_{\left[ \mu\nu\right] }&\simeq&B_{\left[\mu\nu \right]}\otimes \psi\simeq 
\left[ (1,0)\oplus (0,1)\right]\otimes \left[\left(\frac{1}{2},0\right)\oplus 
\left(0, \frac{1}{2} \right) \right]\nonumber\\ 
&\longrightarrow & 
\left[\left(\frac{1}{2},0\right)\oplus \left(0, \frac{1}{2} \right) 
\right]\oplus 
\left[\left( 1,\frac{1}{2}\right)\oplus \left(\frac{1}{2}, 1 \right) \right] 
\oplus  \left[\left( 
\mathbf{ \frac{3}{2} },{\mathbf 0}\right)\oplus \left({\mathbf 0}, 
\mathbf{\frac{3}{2}} \right)\right]. 
\label{tensor_spinor} 
\end{eqnarray} 
Spin-$2$ is part of the antisymmetric tensor-vector 
\begin{eqnarray} 
\Phi_{\left[ \mu\nu\right]\eta }&\simeq&B_{\left[\mu\nu \right]}\otimes 
A_\eta\simeq 
\left[ (1,0)\oplus (0,1)\right]\otimes \left(\frac{1}{2},\frac{1} 
{2}\right)\nonumber\\ 
&\longrightarrow & 
2\left(\frac{1}{2},\frac{1}{2}\right)\oplus 
\left[\left( \mathbf{\frac{1}{2}},\mathbf{ \frac{3}{2}}\right)\oplus 
\left(\mathbf {\frac{3}{2}},\mathbf{ \frac{1}{2}} \right) \right]. 
\label{tensor_vector} 
\end{eqnarray} 
Similarly, spin-$\frac{5}{2}$ can be embedded in the direct product of the 
antisymmetric tensor-vector 
from above with the Dirac spinor, 
giving the totally antisymmetric Lorentz tensor of second rank 
with four-vector-spinor components, $\Psi_{\left[\mu\nu \right]\eta}$, a 
representation space reducible according to 
\begin{eqnarray} 
\Phi_{\left[\mu\nu \right]}\otimes A_\eta \otimes \psi &\simeq&\Psi_{\left[ 
\mu\nu\right]\eta }\simeq 
\left[ (1,0)\oplus (0,1)\right]\otimes 
\left[ \left(\frac{1}{2},\frac{1}{2} \right)\otimes 
\left[\left(\frac{1}{2},0\right)\oplus \left(0, \frac{1}{2} \right) 
\right]\right]\nonumber\\ 
&\longrightarrow & 
2\left[\left(\frac{1}{2},0\right)\oplus \left(0, \frac{1}{2} \right) 
\right]\oplus 
3\left[\left( 1,\frac{1}{2}\right)\oplus \left(\frac{1}{2}, 1 \right) 
\right]\nonumber\\ 
&\oplus& \left[\left(1,\frac{3}{2}\right)\oplus \left(\frac{3}{2}, 1 \right) 
\right] 
\oplus\left[\left({\mathbf 2},\mathbf{\frac{1}{2}}\right)\oplus \left(\mathbf{ 
\frac{1}{2}}, {\mathbf 2} \right) 
\right], 
\label{tensor_4Vspinor} 
\end{eqnarray} 
where the numbers in front of the irreps indicate their multiplicity in the 
reduction scheme.\\ 

\underline{\it Fermion and boson wave equations within the RInS-formalism :}\\ 

\noindent 
In order to exclude the  auxiliary irreducible sectors 
in the above large reducible representation spaces without rising the order of 
the wave equations,  we employ 
static projectors constructed from the Casimir invariants of the Lorentz 
algebra which have the property to unambiguously identify 
anyone of the irreducible representation spaces, no matter whether single- or 
multiple-spin valued  \cite{Wyborne}. 
To be specific, 
\begin{itemize} 
\item \underline{\it Covariant momentum independent  irrep projectors:}\\ 

The Lorentz algebra has two Casimir operators,  denoted by $F$ 
and $G$, and given by \cite{Wyborne} as 
\aec 
~[F]_{AB}&=&\frac{1}{4}[M^{\m\n}]_A{}^C 
[M_{\m\n}]_{C B},\\ 
~[G]_{AB}&=&\frac{1}{4}\e_{\m\n\r\s}[M^{\m\nu}]_A{}^C 
[M^{\r\s}]_{C B} , 
\label{FG_C} 
\cec 
with $A$,$B$, $C$ standing for the generic indexes characterizing the 
representation space of interest. 
  Their respective eigenvalue problems for generic irreducible representation 
spaces of the type $ (j_1,j_2)\oplus (j_2,j_1)$, 
here denoted by $\psi^{(j_1,j_2)}$ are, 

\aec 
F \,\psi^{(j_1,j_2)}&=&\lambda_{(j_1,j_2)}\psi^{(j_1,j_2)},\nonumber\\ 
\lambda_{(j_1,j_2)}&=&\frac{1}{2}(j_1(j_1+1)+j_2(j_2+1))=\frac{1}{2} 
(K(K+2)+M^2)\nonumber\\ 
G \,\psi^{(j_1,j_2)}&=&\eta_{(j_1,j_2)}\psi^{(j_1,j_2)},\nonumber\\ 
\eta_{(j_1,j_2)}&=&\frac{1}{2 i}(j_1(j_1+1)-j_2(j_2+1))=i M(K+1) 
\label{FCasm} 
\cec 
  where 
\aeq 
K=j_1+j_2,\qquad M=\vert j_1-j_2 \vert. 
\ceq 
The $F$ eigenvectors are of well defined parities, while those of 
$G$ are chiral states. In the following we choose to work with the $F$ 
invariant. 

On the basis of  $F$ we design the  following momentum independent Lorentz 
projector, $\mathcal{P}_F^{(j_1,j_2)}$, 

\aec 
\mathcal{P}^{(j_1,j_2)}_F\psi^{(j_1,j_2)} 
=\Pi_{kl}\otimes  \left( 
\frac{ 
F-\l_{(j_k,j_l)} 
}{\l_{(j_1,j_2)}-\l_{(j_k,j_l)}}\right)\psi^{(j_k,j_l)}=\psi^{(j_1,j_2)}, 
\label{feq12} 
\cec 
where $\lambda_{(j_1,j_2)}$ is the eigenvalue of the searched sector, while 
$\lambda_{(j_k,j_l)}$ are the eigenvalues of the auxiliary sectors to 
be excluded. 
The mayor advantage of such projectors is that they are  momentum 
independent and  do not increase the order of the wave equations. 

In what follows we shall consider only such reducible Lorentz tensors 
(or, tensors-spinors) 
which allow the spin-$j$ of our interest to reside within an irreducible 
subspace of maximally two-spins, denoted by $j$ and $j^\prime$, meaning that 
\begin{equation} 
(j_1,j_2)={\Big\{} 
\begin{array}{ccc} 
j_2=\frac{1}{2}, \quad  \mbox{with}&\qquad  j=j_1+\frac{1}{2},&\quad j^\prime 
=j_1-\frac{1}{2},\\ 
j_2=0,  \quad \mbox{with}&j =j_1.& 
\end{array} 
\end{equation} 

\item \underline{{\it Covariant spin projectors, second order in the 
momenta:}} \\ 

The dynamic into the irreducible sector from above carrying the spin of 
interest is then  introduced 
by  applying to it the appropriate  covariant mass-$m$ and spin-$j$ 
projector, 
$\mathcal{P}^{(m,j)}_{{\mathcal W}^2}(p)$, to be occasionally referred  to as 
``Poincar\'e 
projector'', that expresses in terms of 
the  Casimir invariants of the Poincar\'e algebra, the squared four momentum, 
$P^2$, and the squared Pauli-Lubanski vector, ${\mathcal W}^2(p)$, as 
~\cite{Napsuciale:2006wr} 

\aec 
\mathcal{P}^{(m,j)}_{{\mathcal W}^2}(p) \,   \psi^{(m,j)}(p)&= 
&\frac{P^2}{m^2} \(\frac{{\mathcal W}^2(p)-\e_{j^\prime }}{\e_{j}- 
\e_{j^\prime }}\)\, \psi^{(m,j)}(p) 
=\psi^{(m,j)}(p).\label{w2eq12} 
\cec 
Here, ${\mathcal W}^\mu(p)$ denotes the Pauli-Lubanski (pseudo)vector, 
defined as 
\begin{equation} 
\left( {\mathcal W}^{\mu}\right)_{AB}(p)=\frac{1}{2}\epsilon_{\lambda 
\rho\sigma\mu}\left(M^{\rho\sigma} \right)_{ab}p^\mu, 
\label{gen_PL} 
\end{equation} 
where $M^{\rho\sigma}$ are the generators of the Lorentz algebra in the 
representation space of interest, 
while $A$, and $B$ are the sets of indexes that  characterize the 
dimensionality of that very representation space, 
  $\e_{j}=-p^2 j(j+1)$ and $\e_{j^\prime }=-p^2 j^\prime (j^\prime +1)$ are in 
their turn 
the eigenvalues corresponding to the spin-$j$, or, spin- 
$j^\prime$  and mass-$m$  eigenstates of the 
operators ${\mathcal W}^2(p)$, and  $P^2$, respectively. In case of $j_2=0$, 
one sets 
$\e_{j^\prime}=0$. 
In taking this path,  one necessarily  encounters Lagrangians that are second 
order in 
the momenta. 

\noindent 
Second order fermion approaches  have  traditions in field  theory 
\cite{Hostler},\cite{Morgan}, 
and are of growing popularity in QED as well as in QCD \cite{Schubert}, 
\cite{Krasnov}, \cite{DelgadoAcosta:2010nx}. 

\item \underline{\it Product spin-irrep  projectors:}\\ 

Correspondingly, the master equation emerges from combining the covariant 
spin-irrep projectors as 

\begin{eqnarray} 
\left[ \mathcal{P}^{(m,j)}_{{\mathcal W}^2}(p) 
\mathcal{P}^{(j,0)}_F \right]^{\mu_1...\mu_t}_{\nu_1..\nu_t} 
\left[\Psi^{(m,j)}_{(j,0)}(p)\right]_{\mu_1...\mu_t} 
=\left[\Psi^{(m,j)}_{(j,0)}(p)\right]_{\nu_1...\nu_t}, 
\label{masterequations_1} 
\end{eqnarray} 
for pure spin-$j$, or 
\begin{eqnarray} 
\left[   \mathcal{P}^{(m,j)}_{{\mathcal W}^2}(p)\mathcal{P}^{(s,\frac{1} 
{2})}_F 
\right]^{\mu_1...\mu_r}_{\nu_1..\nu_r} 
\left[\Psi^{(m,j)}_{\left(s,\frac{1}{2}\right)}(p)\right]_{\mu_1...\mu_r}= 
\left[\Psi^{(m,j)}_{\left(s,\frac{1}{2}\right)} 
(p)\right]_{\nu_1...\nu_r},\qquad j=s\pm\frac{1}{2}, 
\label{masterequations_2} 
\end{eqnarray} 
for two-spin valued spaces  (\ref{chudo_2}). 

\end{itemize} 

%The method presented here has been applied to the description of the Dirac 
%sector 
%$\left( \frac{1}{2}, 0\right)\oplus \left(0,\frac{1}{2} \right)$ embedded 
%within the Rarita-Schwinger 
%four-vector spinor in a work prior to this \cite{AGK} where 
%it could be shown that it correctly reproduces all the property of a genuine 
%Dirac particle such as 
%the universal $g=2$ value taken by the gyromagnetic ratio, together with the 
%well behaved finite 
%Compton scattering amplitudes in the ultraviolet. 
The present work focuses on the description of the pure spin-$\frac{3}{2}$ 
Weinberg-Joos state, 
$\left(\frac{3}{2}, 0 \right)\oplus \left(0,\frac{3}{2} \right)$ as part of 
the antisymmetric tensor-spinor 
in (\ref{tensor_spinor}).\\ 

{}Though the representation spaces in (\ref{NN_1}) underlying the scheme by 
Niederle and Nikitin are quite different form ours 
in (\ref{chudo_1}), 
occasionally a coincidence can occur, as it indeed happens for pure spin- 
$\frac{3}{2}$ which in both approaches is described in 
terms of the anti-symmetric tensor spinor (\ref{tensor_spinor}). 
However, the assignment in \cite{Niederle}  of spin-$\frac{3}{2}$ to the 
irreducible 
$\left( \frac{3}{2},0\right) \oplus \left(0,\frac{3}{2} \right)$ 
sector of the anti-symmetric tensor spinor of second rank has not been made 
explicit, but might be hidden in the contraction by 
the Dirac matrices, possibly a remnant of a Lorentz projector. 
Also the wave equations in the two methods result different. 
Compared to \cite{Niederle}, our combined Lorentz-and Poincar\'e projector 
method has the advantage to 
apply both to bosons and fermions and to use in general  tensors of lower 
ranks which is 
expected to significantly simplify calculations. 
To be specific, spin-$\frac{5}{2}$ in (\ref{tensor_4Vspinor}) within our 
framework can be 
described by a third rank tensor spinor, while the approach of \cite{Niederle} 
relies upon 
a tensor spinor of fourth rank.\\ \\ 

\section{\label{sec:3}Covariant spin-irrep projectors within  the antisymmetric tensor- 
spinor space} 
%%%%%%%%%%%%%%%%%%%%%%%%%%%%%%%%%%%%%%%%%%%%%%%%%%%%%%%%%%%%%%%%%%%%%%%%%%%%%% 

%%%%%%%%%%%%%%%%%%%%%%%%%%%%%%%%%%%%%%%%% 
%%%%%%%%%%%%%%%%%%%%%%%%%%%%%%%%%%%%%%%%%%%%%%%%%%%%%%%%%%%%%%%%%%%%%%%%%%%%%% 

%%%%%%%%%%%%%%%%%%%%%%%%%%%%%%%%%%%%%%%%% 

The anti-symmetric Lorentz tensor of second rank with spinor components, 
$\Psi_{\left[\mu\nu\right]}$, 
is a two-spin valued  reducible representation space of the $so(1,3)$ algebra 
and has the three 
irreducible sectors $\left(\frac{1}{2},0 \right)\oplus \left( 0, \frac{1} 
{2}\right)$, 
$\left(\frac{1}{2},1 \right)\oplus \left( 1, \frac{1}{2}\right)$, 
and $\left(\frac{3}{2},0 \right)\oplus \left( 0, \frac{3}{2}\right)$ 
  given in the above equation (\ref{tensor_spinor}). 
One  spin-$\frac{1}{2}$ state of say, of positive parity,  resides in the 
Dirac sector, while the other, of negative parity, is part 
of  $\left(\frac{1}{2},1 \right)\oplus \left( 1, \frac{1}{2}\right)$. The 
latter space 
contains in addition a  spin-$\frac{3}{2}$ of same negative parity, opposite 
to the parity of the spin-$\frac{3}{2}$ populating 
the remaining Weinberg-Joos sector, $\left(\frac{3}{2},0 \right)\oplus \left( 
0, \frac{3}{2}\right)$. 

This section starts with the construction of the covariant spin-$\frac{1}{2}$ 
and 
spin-$\frac{3}{2}$  projector operators on the states with mass $m$ within the 
anti-symmetric tensor-spinor space. 
As already announced in the introduction section, the pure-spin $\frac{3}{2}$ 
component of this tensor 
is tracked down by first searching for its spin by means of a  Poincar\'e 
covariant projector constructed along the lines of 
eq.~ (\ref{w2eq12}).  Next, with the aid of eqs.~(\ref{feq12}),  we calculate 
the explicit form of the Lorentz projector 
that localizes the irreducible $\left( \frac{3}{2}, 0\right)\oplus 
\left(0,\frac{3}{2} \right)$ subspace of interest. 
The wave equation is then obtained  executing the prescription of 
(\ref{masterequations_1}). 

%%%%%%%%%%%%%%%%%%%%%%%%%%%%%%%%%%%%%%%%%%%%%%%%%%%%%%%%%%%%%%%%%%%%%%%%%%%%%% 

%%%%%%%%%%%%%%%%%%%%%%%%%%%%%%%%%%%%%%%%% 
\subsection{\label{sec:3.1}The Pauli-Lubanski operator} 
%%%%%%%%%%%%%%%%%%%%%%%%%%%%%%%%%%%%%%%%%%%%%%%%%%%%%%%%%%%%%%%%%%%%%%%%%%%%%% 

%%%%%%%%%%%%%%%%%%%%%%%%%%%%%%%%%%%%%%%%% 

We begin with constructing within the representation space of interest 
the Pauli-Lubanki vector, the key ingredient of the Poincar\'e covariant 
spin-$\frac{3}{2}$ projector. 
The $so(1,3)$ generators within the anti-symmetric tensor-spinor (TS) are 
\aec 
%~[{W^2]_{AB}&=& [W^\m]^A{}_C [W_\m]_{CB}\\ 
%~[W^\l]_{AB}&=&\frac{1}{2}\e^{\l\m\n\s}[M_{\m\n}]_{AB}p_\s,\\ 
~[M^{TS}_{\m\n}]^{ab}_{AB}&=&[M^T_{\m\n}]_{AB}\,\, 
1^{S}_{ab}+\mathbf{1}^T_{AB}\,\, 
\left[M^S_{\m\n}\right]^{ab},\quad A:=\left[\alpha\beta\right], \quad 
B:=\left[\gamma\delta\right],\nonumber\\ 
  M^S_{\m\n}&=&\frac{1}{2}\s_{\m\n}=\frac{i}{4}[\g_\m,\g_\n], 
\label{def:gens} 
\cec 
where the capital letters $A$ and $B$ are the indexes within the $B_{\left[ 
\mu\nu\right]}\sim (1,0)\oplus (0,1)$ tensor (T) part, 
while $a$ and $b$ label the Dirac-spinor (S). In the following, the spinorial 
labels will be suppressed with the aim of 
simplifying notations.  The $[M^{T}_{\m\n}]_{AB}$ generators express in terms 
of the generators, 
$[M^{\left(\frac{1}{2},\frac{1}{2}\right)}_{\m\n}]_{\eta\tau}$ within the 
four-vector, 
$\left(\frac{1}{2},\frac{1}{2}\right)$, 
according to, 
\begin{eqnarray} 
~\left[M_{\m\n}^T\right]_{[\a\b][\g\d]}&=&\frac{1}{2}\left(
\left[M^{\left(\frac{1}{2},\frac{1}{2}\right)}_{\m\n}\right]_{\a\g}g_{\b\d}+
g_{\a\g}\left[M^{\left(\frac{1}{2},\frac{1}{2}\right)}_{\m\n}\right]_{\b\d}\right.\nonumber\\
&&\left.-\left[M_{\m\n}^{\left(\frac{1}{2},\frac{1}{2}\right)}\right]_{\a\d}g_{\b\g}
-g_{\a\d}\left[M_{\m\n}^{\left(\frac{1}{2},\frac{1}{2}\right)}\right]_{\b\g}\right),\\ 
&=&-2\,\,\mathbf{1}_{\a\b}{}^{\k\s}\left[M^{\left(\frac{1}{2},\frac{1} 
{2}\right)}_{\m\n}\right]_{\s} 
{}^\r\mathbf{1}_{\r\k\g\d},\\ 
\mathbf{1}_{\a\b\g\d}&=&\frac{1}{2}(g_{\a\g}g_{\b\d}-g_{\a\d}g_{\b\g}),\\ 
~\left[M^{\left(\frac{1}{2},\frac{1} 
{2}\right)}_{\m\n}\right]_{\a\b}&=&2i\mathbf{1}_{\a\b\m\n}. 
\label{Gen_Tens} 
\end{eqnarray} 
Now using eq.~(\ref{gen_PL}) for the Pauli-Lubanski vectors, 
the Poincar\'e projectors can be obtained  along the line of (\ref{w2eq12}), 
the task of the subsequent subsection. 
%%%%%%%%%%%%%%%%%%%%%%%%%%%%%%%%%%%%%%%%%%%%%%%%%%%%%%%%%%%%%%%%%%%%%%%%%%%%%% 

%%%%%%%%%%%%%%%%%%%%%%%%%%%%%%%%%%%%%%%%% 
\subsection{\label{sec:3.2}The covariant spin-\texorpdfstring{$\frac{1}{2}$}{} and spin- \texorpdfstring{$\frac{3}{2}$}{} 
projectors from the Poincar\'e algebra invariants} 
%%%%%%%%%%%%%%%%%%%%%%%%%%%%%%%%%%%%%%%%%%%%%%%%%%%%%%%%%%%%%%%%%%%%%%%%%%%%%% 

%%%%%%%%%%%%%%%%%%%%%%%%%%%%%%%%%%%%%%%%% 

The tensor-spinor representation space contains only two spin sectors, one 
corresponding to spin $j=\frac{3}{2}$ and the other to $j^\prime =\frac{1} 
{2}$, then the 
corresponding projectors on mass-$m$ and spin-$j$  are found as 
\cite{Napsuciale:2006wr}: 
\aec 
\mathcal{P}_{{\mathcal W}^2}^{\left(m,\frac{1}{2}\right)} 
(p)\psi^{\left(m,\frac{1}{2}\right)}(p)&=&\frac{P^2}{m^2}\ 
\left(\frac{{\mathcal W}^2(p)-\e_{\frac{3}{2}}} 
{\e_{\frac{1}{2}}-\e_{\frac{3}{2}}}\right)\psi^{\left(m,\frac{1}{2}\right)} 
(p)= 
\psi ^{\left(m,\frac{1}{2}\right)}(p)\label{eq:pw212},\\ 
\mathcal{P}_{{\mathcal W}^2}^{(m,\frac{3}{2})}(p)\psi ^{\left(m,\frac{3} 
{2}\right)}(p)&=&\frac{P^2}{m^2}\ 
\left(\frac{{\mathcal W}^2(p)-\e_{\frac{1}{2}}} 
{\e_{\frac{3}{2}}-\e_{\frac{1}{2}}}\right)\psi ^{\left(m,\frac{3}{2}\right)} 
(p)= 
\psi^{\left(m,\frac{3}{2}\right)}(p)\label{eq:pw232}. 
\cec 
Here, $\e_j=-p^2 j(j+1)$ is the ${\mathcal W}^2(p)$ eigenvalue for a generic 
state 
$\psi^{(m,j)}(p)$, 
of mass-$m$ and spin-$j$. We calculate the following explicit expressions, 
\aec 
~\left[\mathcal{P}_{{\mathcal W}^2}^{\left(m,\frac{1}{2}\right)} 
(p)\right]_{[\a\b][\g\d]}&=& 
\frac{2}{m^2} 
(\mathbf{1}_{\a\b\m\s}\mathbf{1}_{\g\d\n\r}+\c_{\a\b\m\s}\c_{\g\d\n\r}) 
p^\s p^\r\left\lbrace\frac{1}{3}\g^\m \g^\n\right\rbrace,\label{eq:pw2x12}\\ 
~\left[\mathcal{P}_{{\mathcal W}^2}^{\left(m,\frac{3}{2}\right)} 
(p)\right]_{[\a\b][\g\d]}&=& 
\frac{2}{m^2} 
(\mathbf{1}_{\a\b\m\s}\mathbf{1}_{\g\d\n\r}+\c_{\a\b\m\s}\c_{\g\d\n\r}) 
p^\s p^\r\left\lbrace g^{\m\n}-\frac{1}{3}\g^\m 
\g^\n\right\rbrace\label{eq:pw2x32}, 
\cec 
where 
\aeq
\c_{\a\b\g\d}=\frac{i}{2}\e_{\a\b\g\d},
\ceq
is the chiral operator in the antisymmetric tensor representation \cite{DelgadoAcosta:2013},
and the terms inside of the  braces recover  the momentum independent 
parts of the well known spin-$\frac{1}{2}$ and spin-$\frac{3}{2}$ projectors 
on the Rarita-Schwinger 
four vector-spinor $(VS)$ space \cite{VanNieuwenhuizen:1981ae} 
\aec 
~\left[\mathbb{P}^{VS}\left(p,\frac{1}{2}\right)\right]_{\a\b}&=&\frac{1} 
{p^2}\(\frac{1} 
{3}\s_{\a\m}\s_{\b\n}\)p^\m p^\n\label{eq:pvs12},\\ 
~\left[\mathbb{P}^{VS}\left(p,\frac{3}{2}\right)\right]_{\a\b}&=&\frac{1} 
{p^2}\(g_{\a\b}g_{\m\n}-\frac{1} 
{3}\s_{\a\m}\s_{\b\n}-g_{\a\m}g_{\b\n}\)p^\m p^\n\label{eq:pvs32}. 
\cec 
%%%%%%%%%%%%%%%%%%%%%%%%%%%%%%%%%%%%%%%%%%%%%%%%%%%%%%%%%%%%%%%%%%%%%%%%%%%%%% 

%%%%%%%%%%%%%%%%%%%%%%%%%%%%%%%%%%%%%%%%% 
\subsection{\label{sec:3.3}Covariant irrep-projectors from the Lorentz algebra invariants } 
%%%%%%%%%%%%%%%%%%%%%%%%%%%%%%%%%%%%%%%%%%%%%%%%%%%%%%%%%%%%%%%%%%%%%%%%%%%%%% 

%%%%%%%%%%%%%%%%%%%%%%%%%%%%%%%%%%%%%%%%% 
The two Casimir invariants, $F$ and $G$  of the Lorentz algebra in 
(\ref{FG_C}) for the antisymmetric tensor- 
spinor are calculated using, 
\aec 
~[F]_{[\a\b][\g\d]}&=&\frac{1}{4}[M^{\m\n}]_{[\a\b]}{}^{[\k\t]} 
[M_{\m\n}]_{[\k\t][\g\d]},\\ 
~[G]_{[\a\b][\g\d]}&=&\frac{1}{4}\e^{\m\n\s\r}[M_{\m\n}]_{[\a\b]}{}^{[\k\t]} 
[M_{\s\r}]_{[\k\t][\g\d]}, 
\cec 
with the generators taken from \eqref{def:gens}.  Their explicit forms are 
obtained as: 
\aec 
~[F]_{[\a\b][\g\d]}&=& 
-\frac{1}{8}\(\s_{\a\b}\s_{\g\d}- 
\s_{\g\d}\s_{\a\b}-22\, \mathbf{1}_{\a\b\g\d}\),\\ 
~[G]_{[\a\b][\g\d]}&=& 
\frac{i}{4}\c_{\a\b}{}^{\s\r}(\s_{\s\r}\s_{\g\d}- 
\s_{\g\d}\s_{\s\r}-16\, \mathbf{1}_{\s\r\g\d})-\frac{3} 
{2}i\g^5\mathbf{1}_{\a\b\g\d}. 
\cec 
The chiral operators $\g^5$ and $\c$ change the parity of the states in the 
respective 
Dirac--and bi-vector sectors of 
representation space considered, so that the eigenstates of the $G$ invariant 
are parity-mixed chiral states. We here choose to work with states of well 
defined parities and 
construct the Lorentz projectors in terms of the  $F$ invariant.\\ 

\noindent 
There are three Lorentz sectors  of the type $(j_2,j_1)\oplus(j_1,j_2)$ in the 
antisymmetric tensor-spinor space, 
corresponding to 
$(j_2,j_1)$ = $\left(\frac{1}{2},0\right)$, $ \left(\frac{1}{2},1\right)$, and 
$\left(\frac{3}{2},0\right)$. 
The associated  generic wave functions, $ \psi^{(j_1,j_2)}$, are 
characterized by their $\l_{(j_1,j_2)}$ eigenvalues with respect to the $F$ 
invariant 
according to: 
\aeq 
F \psi^{(j_2,j_1)}= \l_{(j_1,j_2)} \psi^{(j_2,j_1)}=\frac{1}{2} 
(K(K+2)+M^2)\psi^{(j_1,j_2)}, 
\ceq 
with 
\aeq 
K=j_1+j_2, \qquad M=\vert j_1 -j_2 \vert. 
\ceq 
All three eigenvalues are different and given by, 
\aeq 
\l_{\left(\frac{1}{2},0\right)}=\frac{3}{4},\qquad \l_{\left(\frac{1} 
{2},1\right)}=\frac{11}{4},\qquad 
\l_{\left(\frac{3}{2},0\right)}=\frac{15}{4}. 
\label{tr_org} 
\ceq 
Towards our goal, we  define  operators, 
$\mathcal{Q}^{(j_1^\prime ,j_2^\prime )}$, and 
$\mathcal{Q}^{(j_1^{\prime\prime} ,j_2^{\prime\prime} )}$ that 
suppress those $(j^\prime _1,j_2^\prime)$, and $(j^{\prime\prime} 
_1,j_2^{\prime\prime})$ sectors 
within the tensor under investigation, which are different from the sector we 
are searching for, 
\aec 
\mathcal{Q}^{(j^\prime _1,j^\prime _2)}&=&F-\l_{(j^\prime _1,j^\prime 
_2)}\mathbf{1}. 
\cec 
In effect, the projector onto a selected irreducible Lorentz sector 
$(j_1,j_2)$ can be cast into the form, 
\aeq 
\mathcal{P}^{(j_1,j_2)}_F=\frac{\mathcal{Q}^{(j_1'',j_2'')} 
\mathcal{Q}^{(j_1',j_2')}}{(\l_{(j_1,j_2)}-\l''_{(j_1'',j_2'')}) 
(\l_{(j_1,j_2)}-\l'_{(j_1',j_2')})}, 
\quad j_1^\prime, j_1^{\prime \prime}\not=j_1, \quad j_2^\prime, j_2^{\prime 
\prime}\not=j_1, 
\ceq 
with  $\l_{(j_1,j_2)}$, $\l'_{(j_1',j_2')}$, $\l''_{(j_1'',j_2'')}$ from eq.~ 
(\ref{tr_org}). This way 
we find the following projectors forming a complete set: 
\aec 
~\left[\mathcal{P}^{\left(\frac{1}{2},0\right)}_F\right]_{\a\b\g\d}&=&\frac{1} 
{12}\s_{\a\b}\s_{\g\d}, 
\label{CS_1}\\ 
~\left[\mathcal{P}^{\left(\frac{1} 
{2},1\right)}_F\right]_{\a\b\g\d}&=&\mathbf{1}_{\a\b\g\d}-\frac{1}{8} 
(\s_{\a\b\g\d}+\s_{\g\d}\s_{\a\b}), 
\label{CS_2}\\ 
~\left[\mathcal{P}^{\left(\frac{3}{2},0\right)}_F\right]_{\a\b\g\d}&=&\frac{1} 
{8} 
(\s_{\a\b\g\d}+\s_{\g\d}\s_{\a\b})-\frac{1}{12}\s_{\a\b}\s_{\g\d}. 
\label{CS_3} 
\cec 

%%%%%%%%%%%%%%%%%%%%%%%%%%%%%%%%%%%%%%%%%%%%%%%%%%%%%%%%%%%%%%%%%%%%%%%%%%%%%% 

%%%%%%%%%%%%%%%%%%%%%%%%%%%%%%%%%%%%%%%%% 
\subsection{\label{sec:3.4}Wave equations for particles belonging to  the 
anti-symmetric tensor spinor irreps} 
%%%%%%%%%%%%%%%%%%%%%%%%%%%%%%%%%%%%%%%%%%%%%%%%%%%%%%%%%%%%%%%%%%%%%%%%%%%%%% 

%%%%%%%%%%%%%%%%%%%%%%%%%%%%%%%%%%%%%%%%% 

The commutativity between the $F$ invariant of $so(1,3)$, on the one side, 
and the  $P^2$ and ${\mathcal W}^2(p)$ invariants of 
the Poincar\'e algebra, on the other, makes their diagonalizing  in same 
generic basis 
(here denoted by $\psi^{(m,j)}_{(j_1,j_2)}(p)$) possible. In this manner, the 
spin- 
$j$ of interest 
is unambiguously assigned to the $(j_1,j_2)\oplus (j_2,j_1)$ representation 
space of interest, according to, 
\aeq 
\Pi^{(j_1j_2);j}(p)=\mathcal{P}^{(j_1,j_2)}_F 
\mathcal{P}^{(m,j)}_{{\mathcal W}^2}(p)\, ,\quad 
\Pi^{(j_1j_2);j}(p) 
\Psi^{(m,j)}_{(j_1,j_2)}(p)=\Psi^{(m,j)}_{(j_1,j_2)}(p),\quad j\in \left[|j_1- 
j_2|,(j_1+j_2)\right], 
\label{prpr-eq} 
\ceq 
again, $\Psi^{(m,j)}_{(j_1,j_2)}(p)$  is a generic state of mass $m$, spin $j$ 
that transforms in the 
$(j_1,j_2)$ irrep. 
Within the antisymmetric tensor-spinor there  are six different  products of 
Lorentz and Poincar\'e projectors, 
denoted by $\Pi^{(j_1,j_2); j}(p)$, 
two 
of them vanishing : 
\aec 
\Pi^{\left(\frac{1}{2},0\right);\frac{3}{2}}(p)=\mathcal{P}^{\left(\frac{1} 
{2},0\right)}_F 
\mathcal{P}^{\left(m,\frac{3}{2}\right)}_{{\mathcal 
W}^2}(p)&=&0,\\ 
\Pi^{\left(\frac{3}{2},0\right);\frac{1}{2}}(p)=\mathcal{P}^{\left(\frac{3} 
{2},0\right)}_F \mathcal{P}^{\left(m,\frac{1}{2}\right)}_{{\mathcal 
W}^2}(p)&=&0. 
\cec 
The  remaining four products can be summarized as follows: 
\aeq 
~\left[\Pi^{(j_1,j_2);j}(p)\right]_{[\a\b] 
[\g\d]}\left[\Psi^{(m,j)}_{(j_1,j_2)}\right]^{[\g\d]}(p)=0, \quad j=\frac{1} 
{2}, 
\frac{3}{2}, 
\label{Gl_1} 
\ceq 
where 
\begin{eqnarray} 
~\left[\P^{(j_1,j_2);j}(p)\right]_{[\a\b] 
[\g\d]}&=&\left[\G^{(j_1,j_2);j}_{\m\n}\right]_{[\a\b][\g\d]}p^\m 
p^\n- 
m^2\mathbf{1}_{[\a\b][\g\d]},\\ 
~\left[\G^{(j_1,j_2);j}_{\m\n}\right]_{[\a\b][\g\d]}p^\m p^\n&=&m^2 
[\mathcal{P}^{(j_1,j_2)}_F]_{[\a\b]} 
{}^{[\s\r]}\left[\mathcal{P}^{(m,j)}_{{\mathcal W}^2} 
(p)\right]_{[\s\r][\g\d]}, 
\label{main_eq} 
\end{eqnarray} 
or equivalently 
\aeq 
~[\G^{(j_1,j_2);j}_{\m\n}]_{[\a\b][\g\d]}p^\m p^\n= 
m^2 [\mathcal{P}^{(j_1,j_2)}_F]_{[\a\b]}{}^{[\s\r]} 
[\mathcal{P}^{(m,j)}_{{\mathcal W}^2} 
(p)]_{[\s\r][\h\z]}[\mathcal{P}^{(j_1,j_2)}_F]^{[\h\z]}{}_{[\g\d]}, 
\label{Gl_3} 
\ceq 
explicitly, 
\aec 
~\left[ 
\G^{\left(\frac{1}{2},0\right);\frac{1}{2}}_{\m\n}\right]_{[\a\b] 
[\g\d]}&=&4\left[\mathcal{P}^{\left(\frac{1}{2},0\right)}_F\right]_{[\a\b] 
[\r\m]} 
\frac{1}{3}\g^\r \g^\s \left[\mathcal{P}^{\left(\frac{1} 
{2},0\right)}_F\right]_{[\s\n][\g\d]},\\ 
~\left[\G^{\left(\frac{1}{2},1\right);\frac{1}{2}}_{\m\n}\right]_{[\a\b] 
[\g\d]}&=&4\left[\mathcal{P}^{\left(\frac{1}{2},1\right)}_F\right]_{[\a\b] 
[\r\m]}\ 
\left(\frac{1}{3}\g^\r \g^\s \right) 
\left[\mathcal{P}^{\left(\frac{1}{2},1\right)}_F\right]_{[\s\n][\g\d]},\\ 
~\left[\G^{\left( \frac{1}{2},1\right);\frac{3}{2}}_{\m\n}\right]_{[\a\b] 
[\g\d]}&=&4\left[\mathcal{P}^{\left(\frac{1}{2},1\right)}_F\right]_{[\a\b] 
[\r\m]}\ 
\left(g^{\r\s}-\frac{1}{3}\g^\r \g^\s \right) 
\left[\mathcal{P}^{\left(\frac{1}{2},1\right)}_F\right]_{[\s\n][\g\d]},\\ 
~\left[\G^{\left(\frac{3}{2},0\right);\frac{3}{2})}_{\m\n}\right]_{[\a\b] 
[\g\d]}&=&4\left[\mathcal{P}^{\left(\frac{3}{2},0\right)}_F\right]_{[\a\b] 
[\r\m]}\ 
\left(g^{\r\s}-\frac{1}{3}\g^\r \g^\s \right) 
\left[\mathcal{P}^{\left(\frac{3}{2},0\right)}_F\right]_{[\s\n] 
[\g\d]}\label{gamma-s32}. 
\cec 
The propagators are found by inverting (\ref{main_eq}) as, 
\aec\label{eq:props} 
~\left[{\mathcal S}^{(j_1,j_2);j}(p)\right]_{[\a\b][\g\d]}&=&\left( 
\left[\Pi^{(j_1,j_2);j}(p)\right]_{[\a\b][\g\d]}-m^2\mathbf{1}_{[\a\b][\g\d]} 
\right)^{-1}\\ 
&=&\frac{\left[\D^{(j_1,j_2);j}(p)\right]_{[\a\b][\g\d]}}{p^2-m^2},\\ 
~\left[\D^{(j_1,j_2);j}(p)\right]_{[\a\b][\g\d]}&=&\frac{1} 
{m^2}\left[\G_{\m\nu}^{(j_1,j_2);j} 
\right]_{[\a\b][\g\d]}p^\m p^\n-\frac{(p^2-m^2)}{m^2}\mathbf{1}_{[\a\b] 
[\g\d]}. 
\cec 
%%%%%%%%%%%%%%%%%%%%%%%%%%%%%%%%%%%%%%%%%%%%%%%%%%%%%%%%%%%%%%%%%%%%%%%%%%%%%% 

%%%%%%%%%%%%%%%%%%%%%%%%%%%%%%%%%%%%%%%%% 
%%%%%%%%%%%%%%%%%%%%%%%%%%%%%%%%%%%%%%%%%%%%%%%%%%%%%%%%%%%%%%%%%%%%%%%%%%%%%% 

%%%%%%%%%%%%%%%%%%%%%%%%%%%%%%%%%%%%%%%%% 
\section{\label{sec:4}The explicit irreducible degrees of freedom spanning the 
antisymmetric tensor spinor space} 
This section is devoted to the explicit construction of the 24 irreducible 
degrees of 
freedom spanning the antisymmetric tensor spinor space. 
In what follows we shall systematically omit the brackets indicating the 
antisymmetric indexes for the purpose of simplifying notations 
and hope that this will not create confusions. 

%%%%%%%%%%%%%%%%%%%%%%%%%%%%%%%%%%%%%%%%%%%%%%%%%%%%%%%%%%%%%%%%%%%%%%%%%%%%%% 

%%%%%%%%%%%%%%%%%%%%%%%%%%%%%%%%%%%%%%%%% 
%%%%%%%%%%%%%%%%%%%%%%%%%%%%%%%%%%%%%%%%%%%%%%%%%%%%%%%%%%%%%%%%%%%%%%%%%%%%%% 

%%%%%%%%%%%%%%%%%%%%%%%%%%%%%%%%%%%%%%%%% 

%%%%%%%%%%%%%%%%%%%%%%%%%%%%%%%%%%%%%%%%%%%%%%%%%%%%%%%%%%%%%%%%%%%%%%%%%%%%%% 

%%%%%%%%%%%%%%%%%%%%%%%%%%%%%%%%%%%%%%%%% 
\subsection{\label{sec:4.1}Diagonalizing the  covariant spin  projector} 
%%%%%%%%%%%%%%%%%%%%%%%%%%%%%%%%%%%%%%%%%%%%%%%%%%%%%%%%%%%%%%%%%%%%%%%%%%%%%% 

%%%%%%%%%%%%%%%%%%%%%%%%%%%%%%%%%%%%%%%%% 

%----------------------------------------------------------------------------- 
\subsubsection{\label{sec:4.1.1}Spin-\texorpdfstring{$\frac{1}{2}$}{} states within the antisymmetric tensor 
spinor and  Dirac spinors} 
%-------------------------------------- 
The direct-product nature of the antisymmetric tensor spinor 
(\ref{tensor_spinor}) allows to construct the spin states residing there 
making use of angular momentum addition theorems. There are two 
sorts of spin $ \frac{1}{2}$ states one can construct, the first kind is built 
up from 
  positive parity spin-$1^+$ vectors spanning   $(1,0)\oplus(0,1)$, here 
  denoted by $[\h_{+}(\mathbf{p},\ell)]^{\a\b}$,  and Dirac's $u$ and $v$ 
spinors of opposite parities, here denoted as 
$u_{\pm}(\mathbf{p},\l)$. The resulting spin-$\frac{1}{2}$ Clebsh-Gordan 
combination being 
$\left[U^{(+)}_{\pm}\left(\mathbf{p},\frac{1}{2},\l\right)\right]^{\a\b}$: 
\aec 
~\left[U^{(+)}_{\pm}\left(\mathbf{p},\frac{1}{2},\frac{1} 
{2}\right)\right]^{\a\b}&=&-\sqrt{\frac{1}{3}}[\h_{+} 
(\mathbf{p},0)]^{\a\b}u_{\pm}\left(\mathbf{p},\frac{1}{2}\right) 
+\sqrt{\frac{2}{3}}[\h_{+}(\mathbf{p},1)]^{\a\b}u_{\pm}\left(\mathbf{p},- 
\frac{1}{2}\right),\\ 
~\left[U^{(+)}_{\pm}\left(\mathbf{p},\frac{1}{2},-\frac{1} 
{2}\right)\right]^{\a\b}&=&\sqrt{\frac{1}{3}}[\h_{+} 
(\mathbf{p},0)]^{\a\b}u_{\pm}\left(\mathbf{p},-\frac{1}{2}\right) 
-\sqrt{\frac{2}{3}}[\h_{+} 
(\mathbf{p},-1)]^{\a\b}u_{\pm}\left(\mathbf{p},\frac{1}{2}\right),\\ 
u({\mathbf p},\l)=u_+({\mathbf p},\l), &\quad& v({\mathbf p},\l)=u_-({\mathbf 
p},\l). 
\label{onehpl} 
\cec 
The second kind spinors refer to the coupling between the  negative parity 
spin-$1^-$ vectors from 
$(1,0)\oplus (0,1)$, denoted by $[\h_{-}(\mathbf{p},\ell)]^{\a\b}$, and same 
Dirac spinors from above, giving: 
\aec 
~\left[U^{(-)}_{\pm}\left(\mathbf{p},\frac{1}{2},\frac{1} 
{2}\right)\right]^{\a\b}&=&-\sqrt{\frac{1}{3}}[\h_{-} 
(\mathbf{p},0)]^{\a\b}u_{\mp}\left(\mathbf{p},\frac{1}{2}\right) 
+\sqrt{\frac{2}{3}}[\h_{-}(\mathbf{p},1)]^{\a\b}u_{\mp}\left(\mathbf{p},- 
\frac{1}{2}\right),\\ 
~\left[U^{(-)}_{\pm}\left(\mathbf{p},\frac{1}{2},-\frac{1} 
{2}\right)\right]^{\a\b}&=&\sqrt{\frac{1}{3}}[\h_{-} 
(\mathbf{p},0)]^{\a\b}u_{\mp}\left(\mathbf{p},-\frac{1}{2}\right) 
-\sqrt{\frac{2}{3}}[\h_{-} 
(\mathbf{p},-1)]^{\a\b}u_{\mp}\left(\mathbf{p},\frac{1}{2}\right). 
\label{onehmi} 
\cec 
There is a remarkable relationship between the basis states, 
$A^\mu_\pm ({\mathbf p},\ell)$,  within the four-vector space 
(with the low case signs denoting the parity) , and the associated Pauli- 
Lubanski vector, $W^{\left(\frac{1}{2},\frac{1}{2}\right)}_\m(p)$, on the one 
side, 
and the basis states within the anti-symmetric tensor spinor, on the other. 
Namely, these bases are intertwined 
  as follows, 
\aec 
~[\h_{+}(\mathbf{p},\ell)]^{\a\b}&=&-\frac{1}{\sqrt{2}m} 
\left[W^{\left(\frac{1}{2},\frac{1}{2}\right)}_\m(p)\right]^{\a\b}A^\m_- 
(\mathbf{p},\ell),\label{rel:tw}\\ 
~[\h_{-}(\mathbf{p},\ell)]^{\a\b}&=&\c^{\a\b}{}_{\g\d} [\h_{+} 
(\mathbf{p},\ell)]^{\g\d}, 
\label{rel:dt} 
\cec 
where $W^{\left(\frac{1}{2},\frac{1}{2}\right)}_\m(p)$ is, 
\aeq 
~\left[W^{\left(\frac{1}{2},\frac{1}{2}\right)}\, ^\l 
(p)\right]^{\a\b}=\frac{1}{2}\e^{\l\s\r\m} 
[M^{\left(\frac{1}{2},\frac{1}{2}\right)}_{\s\r}]^{\a\b}p_\m. 
\ceq 
In result,  the eqs.~(\ref{onehpl}) and (\ref{onehmi}) simplify to, 
\aeq 
~\left[U^{(+)}_{\pm}\left(\mathbf{p},\frac{1}{2},\l \right)\right]^{\a\b}=- 
\frac{1}{\sqrt{2}m} 
[W^{\left(\frac{1}{2},\frac{1}{2}\right)}_\m(p)]^{\a\b}\g^5 {\mathcal 
U}^\mu_\pm\left(\mathbf{p},\frac{1}{2},\l\right). 
\ceq 
Here, $A^\a_\pm({\mathbf p},\ell)$ and $u_{\pm}({\mathbf p},\l )$ have now 
been absorbed by the spin-$\frac{1}{2}$ 
Rarita-Schwinger  four-vector-spinor ${\mathcal U}^\mu _\pm\left({\mathbf 
p},\frac{1}{2},\l\right)$. A 
further significant simplification is achieved by noticing that 
  spin-$\frac{1}{2}$ four-vector spinors, ${\mathcal U}^\mu _\pm\left({\mathbf 
p},\frac{1}{2},\l\right)$, by 
themselves   can be re-expressed in terms of the Dirac spinor and the Pauli- 
Lubanski vector, $w^\a (p)$,  in 
$\left(\frac{1}{2},0\right)\oplus\left( 0,\frac{1}{2}\right)$. Namely, the 
following relation holds valid, 
\begin{eqnarray} 
~{\mathcal U}^\a_\pm\left(\mathbf{p},\frac{1}{2},\l\right)&=& 
\frac{2}{\sqrt{3}m}[\omega (p)]^\a \g^5 u_{\pm} 
(\mathbf{p},\l).\label{eq:vsus}\\ 
~[\omega (p)]^\a &=& -\frac{i}{2}\g^5\s^{\a\n}p_\n\label{PL_Dir}. 
\end{eqnarray} 
In effect, the spin-$\frac{1}{2}$ degrees of freedom within the antisymmetric 
tensor spinor 
equivalently rewrite as, 
\aec 
~\left[U^{(+)}_{\pm}\left(\mathbf{p},\frac{1}{2},\l\right)\right]^{\a\b}&=&- 
\frac{2}{\sqrt{6}m^2} 
[W^{\left(\frac{1}{2},\frac{1}{2}\right)}_\m(p)]^{\a\b}\g^5\omega^\m(p) \g^5 
u_{\pm}(\mathbf{p},\l),\\ 
~\left[U^{(-)}_{\pm}\left(\mathbf{p},\frac{1}{2},\l\right)\right]^{\a\b}&=&- 
\frac{2}{\sqrt{6}m^2}\c^{\a\b} 
{}_{\g\d}[W^{\left(\frac{1}{2},\frac{1}{2}\right)}_\m(p)]^{\g\d}\g^5\omega^\m 
(p)]u_{\pm}(\mathbf{p},\l), 
\cec 
yielding the following compact expressions: 
\aec 
~\left[U^{(+)}_{\pm}\left(\mathbf{p},\frac{1} 
{2},\l\right)\right]^{\a\b}&=&\frac{2} 
{\sqrt{6}m^2}\c^{\a\b} 
{}_{\s\r} p^\s\g^\r \slashed{p}\g^5 u_{\pm}(\mathbf{p},\l),\\ 
~\left[U^{(-)}_{\pm}\left(\mathbf{p},\frac{1} 
{2},\l\right)\right]^{\a\b}&=&\frac{2} 
{\sqrt{6}m^2}\mathbf{1}^{\a\b}{}_{\s\r} p^\s\g^\r \slashed{p} u_{\pm} 
(\mathbf{p},\l). 
\cec 
The conjugate states are now easy to define in terms of the Dirac conjugate 
spinors and read, 
\aeq 
~\left[\overline{U}^{(+)/(-)}_{\pm} 
\left(\mathbf{p},\frac{1}{2},\l\right)\right]^{\a\b}= 
\left( 
\left[\g^0 U^{(+)/(-)}_{\pm} 
\left(\mathbf{p},\frac{1}{2},\l\right)\right]^{\a\b}\right)^{\dagger}. 
\ceq 
The explicit expressions are, 
\aec 
~\left[\overline{U}^{(+)}_{\pm}\left(\mathbf{p},\frac{1} 
{2},\l\right)\right]^{\a\b}&=&\frac{2} 
{\sqrt{6}m^2}\overline{u}_{\pm}(\mathbf{p},\l)\g^5\slashed{p}\g^\r 
p^\s\c_{\s\r}{}^{\a\b},\\ 
~\left[\overline{U}^{(-)}_{\pm}\left(\mathbf{p},\frac{1} 
{2},\l\right)\right]^{\a\b}&=&\frac{2} 
{\sqrt{6}m^2}\overline{u}_{\pm}(\mathbf{p},\l)\slashed{p}\g^\r 
p^\s\mathbf{1}_{\s\r}{}^{\a\b} , 
\cec 
where we have used $(\c^{\a\b\g\d})^*=-\c^{\a\b\g\d}$ and $\g^5\g^0=-\g^0 
\g^5$. The spin-$\frac{1}{2}$ states are normalized as, 
\aeq 
~\left[\overline{U}^{(+)/(-)}_{\pm}\left(\mathbf{p},\frac{1} 
{2},\l'\right)\right]^{\a\b} 
\left[U^{(+)/(-)}_{\pm}\left(\mathbf{p},\frac{1} 
{2},\l\right)\right]_{\a\b}=\pm 1\d^{\l'}_\l, 
\ceq 
and orthogonal according to, 
\aec 
~\left[\overline{U}^{(+)}_{\pm}\left(\mathbf{p},\frac{1} 
{2},\l'\right)\right]^{\a\b} \left[U^{(- 
)}_{\pm} 
\left(\mathbf{p},\frac{1}{2},\l\right)\right]_{\a\b}=0,\\ 
~\left[\overline{U}^{(-)}_{\pm}\left(\mathbf{p},\frac{1} 
{2},\l'\right)\right]^{\a\b} 
\left[U^{(+)}_{\pm} 
\left(\mathbf{p},\frac{1}{2},\l\right)\right]_{\a\b}=0. 
\cec 
%--------------------------------------------------------- 
\subsubsection{\label{sec:4.1.2}Spin-\texorpdfstring{$\frac{3}{2}$}{} states within the antisymmetric tensor 
spinor and Rarita-Schwinger four-vector spinors} 
%---------------------------------------------- 
The spin-$\frac{3}{2}$ states arising from the coupling of the  positive 
parity vectors spanning the 
($1,0)\oplus (0,1)$ space with the Dirac $u$ and $v$ spinors (here denoted by 
$u_\pm$) are 
\aec 
~\left[U_{\pm}^{(+)}\left(\mathbf{p},\frac{3}{2},\frac{3} 
{2}\right)\right]^{\a\b}&=&[\h_+ 
(\mathbf{p},1)]^{\a\b}u_\pm \left(\mathbf{p},\frac{1}{2}\right),\\ 
~\left[U_{\pm}^{(+)}\left(\mathbf{p},\frac{3}{2},\frac{1} 
{2}\right)\right]^{\a\b}&=&\frac{1}{\sqrt{3}} 
[\h_+ 
(\mathbf{p},1)]^{\a\b}u_{\pm}\left(\mathbf{p},-\frac{1}{2}\right) 
+\sqrt{\frac{2}{3}}\left[\h_+(\mathbf{p},0)\right]^{\a\b}u_{\pm} 
\left(\mathbf{p},\frac{1}{2}\right),\\ 
~\left[U_{\pm}^{(+)}\left(\mathbf{p},\frac{3}{2},-\frac{1} 
{2}\right)\right]^{\a\b}&=&\frac{1}{\sqrt{3}} 
[\h_+ 
(\mathbf{p},-1)]^{\a\b}u_{\pm}\left(\mathbf{p},\frac{1}{2}\right) 
+\sqrt{\frac{2}{3}}[\h_+(\mathbf{p},0)]^{\a\b}u_{\pm}\left(\mathbf{p},- 
\frac{1}{2}\right),\\ 
~\left[U_{\pm}^{(+)}\left(\mathbf{p},\frac{3}{2},-\frac{3} 
{2}\right)\right]^{\a\b}&=&[\h_+ 
(\mathbf{p},-1)]^{\a\b}u_{\pm}\left(\mathbf{p},-\frac{1}{2}\right). 
\cec 
Again, making use of \eqref{rel:tw}, allows for the simplifications, 
\aeq 
~\left[U_{\pm}^{(+)}\left(\mathbf{p},\frac{3}{2},\l\right)\right]^{\a\b}=- 
\frac{1}{\sqrt{2}m} 
[W^{\left(\frac{1}{2},\frac{1}{2}\right)}_\m(p)]^{\a\b}\g^5{\mathcal 
U}^\mu_\pm 
\left(\mathbf{p},\frac{3}{2},\l\right), 
\ceq 
where ${\mathcal U}_\pm^\mu\left(\mathbf{p},\frac{3}{2},\l\right)$ are the 
standard Rarita-Schwinger  spin- 
$\frac{3}{2}$  four vector-spinors. 
Explicit use of the expression of $M^{\left(\frac{1}{2},\frac{1} 
{2}\right)}_{\mu\nu}$ in (\ref{Gen_Tens}) to 
calculate $W^{\left(\frac{1}{2},\frac{1}{2}\right)}_\mu(p)$ amounts again to 
a relationship between the spin-$\frac{3}{2}$ degrees within the anti- 
symmetric tensor spinor and those within the 
four-vector spinor, 
\aec 
~\left[U_{\pm}^{(+)}\left (\mathbf{p},\frac{3}{2},\l\right)\right]^{\a\b}&=&- 
\frac{2} 
{\sqrt{2}m}\c^{\a\b} 
{}_{\m\n}\g^5 {\mathcal U}^\m_\pm\left(\mathbf{p},\frac{3}{2},\l\right) 
p^\n,\\ 
~\left[U_{\pm}^{(-)}\left(\mathbf{p},\frac{3}{2},\l\right)\right]^{\a\b}&=&- 
\frac{2} 
{\sqrt{2}m}\mathbf{1}^{\a\b}{}_{\m\n}{\mathcal 
U}^\m_\pm\left(\mathbf{p},\frac{3}{2},\l\right)p^\n. 
\cec 
{}Finally,  the couplings of the spin-$1^-$ vectors from $(1,0)\oplus(0,1)$ 
with the Dirac spinors emerge as 
\aeq 
~\left[U_{\pm}^{(-)}\left(\mathbf{p},\frac{3} 
{2},\l\right)\right]^{\a\b}=\c^{\a\b} 
{}_{\g\d}\left[U_{\pm}^{(+)} 
\left(\mathbf{p},\frac{3}{2},\l\right)\right]^{\g\d}. 
\ceq 
The respective  conjugate states are then found according to, 
\aec 
~\left[\overline{U}_{\pm}^{(+)}\left(\mathbf{p},\frac{3} 
{2},\l\right)\right]^{\a\b}&=&-\frac{2} 
{\sqrt{2}m} 
\overline{{\mathcal U}}^\m_\pm\left(\mathbf{p},\frac{3}{2},\l\right)\g^5 
p^\n\c_{\m\n}{}^{\a\b} ,\\ 
~\left[\overline{U}_{\pm}^{(-)}\left(\mathbf{p},\frac{3} 
{2},\l\right)\right]^{\a\b}&=&-\frac{2} 
{\sqrt{2}m}\mathbf{1}^{\a\b}{}_{\m\n}{\mathcal U}^\m_\pm 
\left(\mathbf{p},\frac{3}{2},\l\right)p^\n. 
\cec 
The latter expressions allow for an easy calculation of the norms of the 
states under discussion as 

\aeq 
~\left [\overline{U}^{(+)/(-)}_{\pm}\left(\mathbf{p},\frac{3} 
{2},\l'\right)\right]^{\a\b} 
\left[U^{(+)/(-)}_{\pm} 
\left(\mathbf{p},\frac{3}{2},\l\right)\right]_{\a\b}=\pm {\bf 1}\d^{\l'}_\l, 
\ceq 
and obey the following relationships, 
\aec 
~\left[\overline{U}^{(+)}_{\pm}\left(\mathbf{p},\frac{3} 
{2},\l'\right)\right]^{\a\b} 
\left[U^{(-)}_{\pm}\left(\mathbf{p},\frac{3}{2},\l\right)\right]_{\a\b}=0,\\ 
~\left[\overline{U}^{(-)}_{\pm}\left(\mathbf{p},\frac{3} 
{2},\l'\right)\right]^{\a\b} 
\left[U^{(+)}_{\pm} 
\left(\mathbf{p},\frac{3}{2},\l\right)\right]_{\a\b}=0. 
\cec 
%----------------------------------------------------------------------------- 

%%%%%%%%%%%%%%%%%%%%%%%%%%%%%%%%%%%%%%%%%%%%%%%%%%%%%%%%%%%%%%%%%%%%%%%%%%%%%% 

%%%%%%%%%%%%%%%%%%%%%%%%%%%%%%%%%%%%%%%%% 
\subsection{\label{sec:4.2}Diagonalizing the covariant spin--irrep  projectors } 

%%%%%%%%%%%%%%%%%%%%%%%%%%%%%%%%%%%%%%%%%%%%%%%%%%%%%%%%%%%%%%%%%%%%%%%%%%%%%% 

%%%%%%%%%%%%%%%%%%%%%%%%%%%%%%%%%%%%%%%%% 
It is not difficult to verify that none of the sets of $U$ states diagonalizing a 
covariant spin projector 
is an eigenstate to the  invariants of the Lorentz algebra from the above 
subsection \ref{sec:3.3}. 
The spin-$\frac{1}{2}$ states that diagonalize the $F$ invariant, 
here denoted by 
$w_\pm ^{\left(\frac{1}{2},0\right)}\left({\mathbf p},\frac{1}{2},\l \right)$, 
with the lower case index,  $\pm$,  again indicating the parity, 
are found, modulo a 
constant, through the projections 
of the spin-$\frac{1}{2}$ states diagonalizing the Poincar\'e projector from 
the previous subsection  on 
$\left( \frac{1}{2},0 \right)\oplus \left( 0, \frac{1}{2}\right)$ 
as following, 

\aec 
~\left[w^{\left(\frac{1}{2},0\right)}_{\pm}\left(\mathbf{p},\frac{1} 
{2},\l\right)\right]^{\a\b}&=&N 
\left[\mathcal{P}_F^{\left(\frac{1}{2},0\right)}\right]^{\a\b\g\d}\left[U^{(- 
)}_{\pm} 
\left(\mathbf{p},\frac{1}{2},\l\right)\right]_{\g\d},\\ 
&=&N\left[\mathcal{P}_F^{\left(\frac{1} 
{2},0\right)}\right]^{\a\b\g\d}\left[U^{(+)}_{\pm} 
\left(\mathbf{p},\frac{1}{2},\l\right)\right]_{\g\d}, 
\label{iird32} 
\cec 
where $N$ is a normalization factor and we have used 
\aeq 
~\left[\mathcal{P}_F^{\left(\frac{1} 
{2},0\right)}\right]^{\a\b\r\s}\c_{\r\s\g\d}\g^5=~ 
\left[\mathcal{P}_F^{\left(\frac{1}{2},0\right)}\right]^{\a\b}{}_{\g\d}= 
\left[\mathcal{P}_F^{\left(\frac{1} 
{2},0\right)}\right]^{\a\b\r\s}\mathbf{1}_{\r\s\g\d}. 
\ceq 
In a way similar, the irreducible states residing within 
$\left(\frac{1}{2},1\right)\oplus\left(1,\frac{1}{2}\right)$ are defined 
according to, 
\aec 
~\left[w^{\left(\frac{1}{2},1\right)}_{\pm}(\mathbf{p},j,\l)\right]^{\a\b}&=&- 
N\left[\mathcal{P}_F^{\left(\frac{1}{2},1\right)}\right]^{\a\b\g\d}\left[U^{(- 
)}_{\pm} 
(\mathbf{p},j,\l)\right]_{\g\d},\\ 
&=&N\left[\mathcal{P}_F^{\left(\frac{1} 
{2},1\right)}\right]^{\a\b\g\d}\left[U^{(+)}_{\pm} 
(\mathbf{p},j,\l)\right]_{\g\d}, 
\cec 
with $j=\frac{1}{2},\frac{3}{2}$. Further use has been made of the following 
relationships, 
\aeq 
~-[\mathcal{P}_F^{\left(\frac{1}{2},1\right)}]^{\a\b\r\s}\c_{\r\s\g\d}\g^5=~ 
[\mathcal{P}_F^{\left(\frac{1}{2},1\right)}]^{\a\b}{}_{\g\d}= 
[\mathcal{P}_F^{\left(\frac{1}{2},1\right)}]^{\a\b\r\s}\mathbf{1}_{\r\s\g\d}. 
\label{relship1} 
\ceq 
{}Finally for the last sector $\left(\frac{3}{2},0\right)\oplus\left(0,\frac{3}{2}\right)$ we have, 
\aec 
~\left[w^{\left(\frac{3}{2},0\right)}_{\pm} 
\left(\mathbf{p},\frac{3}{2},\l\right)\right]^{\a\b}&=&N 
\left[\mathcal{P}_F^{\left(\frac{3}{2},0\right)}\right]^{\a\b\g\d} 
\left[U^{(-)}_{\pm}\left(\mathbf{p},\frac{3}{2},\l\right)\right]_{\g\d},\\ 
&=&N\left[\mathcal{P}_F^{\left(\frac{3}{2},0\right)}\right]^{\a\b\g\d} 
\left[U^{(+)}_{\pm} 
\left(\mathbf{p},\frac{3}{2},\l\right)\right]_{\g\d}. 
\cec 
Here, use has been made of a relationship similar to that in (\ref{relship1}) 
and given by, 
\aeq 
~\left[\mathcal{P}_F^{\left(\frac{3} 
{2},0\right)}\right]^{\a\b\r\s}\c_{\r\s\g\d}\g^5=~ 
\left[\mathcal{P}_F^{\left(\frac{3}{2},0\right)}\right]^{\a\b}{}_{\g\d}= 
\left[\mathcal{P}_F^{\left(\frac{3} 
{2},0\right)}\right]^{\a\b\r\s}\mathbf{1}_{\r\s\g\d}. 
\label{relship2} 
\ceq 
In result, the states under discussion are related to  the Rarita-Schwinger 
four vector-spinors, ${\mathcal U}_\pm^\a (\mathbf{p},j,\l)$, as 
\aec 
~[w^{(j_1,j_2)}_{\pm} 
(\mathbf{p},j,\l)]^{\a\b}&=&N[\mathcal{P}_F^{(j_1,j_2)}]^{\a\b\g\d} 
\left[U^{(+)}_{\pm}(\mathbf{p},j,\l)\right]_{\g\d}\\ 
&=&\frac{2 N}{\sqrt{2}m}[\mathcal{P}_F^{(j_1,j_2)}]^{\a\b}{}_{\m\n}p^\m 
{\mathcal U}^\n_{\pm}(\mathbf{p},j,\l). 
\cec 

The constant factor $N$ has to be chosen in a way ensuring the normalization 
of these states to $\pm 1$ 
in dependence on their respective parity, positive versus negative. 
Such a normalization is guaranteed by  $N=\sqrt{2}$. 
In effect,  all the irreducible degrees of freedom within the antisymmetric 
tensor-spinor 
can be expressed  in terms of four-vector spinors according to. 

\aeq 
~[w^{(j_1,j_2)}_{\pm}(\mathbf{p},j,\l)]^{\a\b}=[f^{(j_1,j_2)} 
(\mathbf{p})]^{\a\b\m}{\mathcal U}_\m \, _{(\pm)}(\mathbf{p},j,\l) 
\label{eq:atsvs} 
\ceq 
with the $~[f^{(j_1,j_2)}{(p)}]^{\a\b\m}$  tensors being defined as 
\aeq 
~[f^{(j_1,j_2)}{(p)}]^{\a\b\m}=\frac{2}{m} 
[\mathcal{P}_F^{(j_1,j_2)}]^{\a\b\g\m}p_\g, 
\label{f_tensors} 
\ceq 
and the projectors form eqs.~(\ref{CS_1})-(\ref{CS_3}). 
The explicit expressions for $~[f^{(j_1,j_2)}{(p)}]^{\a\b\m}$ are, 
\aec 
~\left[f^{\left(\frac{1}{2},0\right)}(\mathbf{p})\right]^{\a\b\m}&=&\frac{1} 
{6m}\s^{\a\b}\s^{\g\m}p_\g, 
\label{f1}\\ 
~\left[f^{\left(\frac{1}{2},1\right)}(\mathbf{p})\right]^{\a\b\m}&=&\frac{2} 
{m}\mathbf{1}^{\a\b\g\m}p_\g- 
\frac{1}{4 m}(\s^{\a\b}\s^{\g\m}+\s^{\g\m}\s^{\a\b})p_\g, 
\label{f2}\\ 
~\left[f^{\left(\frac{3}{2},0\right)}(\mathbf{p})\right]^{\a\b\m}&=&\frac{1}{4 
m} 
(\s^{\a\b}\s^{\g\m}+\s^{\g\m}\s^{\a\b})p_\g-\frac{1} 
{6m}\s^{\a\b}\s^{\g\m}p_\g. 
\label{f3} 
\cec 

The orthogonality of the  Lorentz projectors implies orthogonality of  the 
$f$-tensors 
according to, 
\aeq 
~[\overline{f}^{(j_1',j_2')}(p)]^{\a\b\m}[f^{(j_1,j_2)}(p)]_{\a\b\m}=0, 
\ceq 
for $(j_1',j_2')\neq(j_1,j_2)$, and $[\overline{f}^{(j_1,j_2)} 
(p)]^{\a\b\m}=\g^0 ([f^{(j_1,j_2)}(p)]^{\a\b\m})^\dagger\g^0$. 
{}For  $(j_1',j_2')=(j_1,j_2)$ the above tensors obey the relations 
\aeq 
~[f^{(j_1,j_2)}(p)]^{\a\b\m}[ \overline{f}^{(j_1,j_2)}(p)]^{\g\d} 
{}_{\m}=\frac{p^2}{m^2} 
[\mathcal{P}_F^{(j_1,j_2)}]^{\a\b\g\d}\label{rel:fsouter}, 
\ceq 
and 
\aec 
~\left[\overline{f}^{\left(\frac{1}{2},0\right)}(p)\right]^{\a\b}{}_{\m} 
[f^{\left(\frac{1}{2},0\right)}(p)]_{\a\b\n}&=&\frac{1} 
{m^2} \left(\frac{1}{3}\s_{\a\m}\s_{\b\n}\right)p^\a 
p^\b,\label{eq:fintern12}\\ 
~\left[\overline{f}^{\left(\frac{1}{2},1\right)}(p)\right]^{\a\b}{}_{\m} 
\left[f^{\left(\frac{1}{2},1\right)}(p)\right]_{\a\b\n}&=&\frac{1} 
{m^2}  (g_{\a\b}g_{\m\n}-g_{\a\m}g_{\b\n})p^\a p^\b,\label{eq:fintern121}\\ 
~\left[\overline{f}^{\left(\frac{3}{2},0\right)}(p)\right]^{\a\b}{}_{\m} 
\left[f^{\left(\frac{3}{2},0\right)}(p)\right]_{\a\b\n}&=&\frac{1} 
{m^2}  (g_{\a\b}g_{\m\n}-\frac{1}{3}\s_{\a\m}\s_{\b\n}-g_{\a\m}g_{\b\n})p^\a 
p^\b. 
\label{eq:fintern320} 
\cec 
The proof of the normalization to $(\pm1)$ of the Lorentz eigenstates follows 
upon recognizing here, the first product as the spin-$\frac{1}{2}$ projector 
over the vector- 
spinors, ${\mathcal U}^\mu_{\pm}\left(\mathbf{p},\frac{1}{2},\l\right)$, the 
second product as the spin-$1^-$ 
projector over the four vectors $A^\mu_{-}(\mathbf{p},\ell)$, and the third 
product as the spin-$\frac{3}{2}$ projector over the four vector-spinors, 
${\mathcal U}_{\pm}^\mu \left(\mathbf{p},\frac{3}{2},\l\right)$, thus ending 
up with 
\aec 
~[\overline{w}^{(j_1,j_2)}_{\pm}(\mathbf{p},j,\l)]^{\a\b}[w^{(j_1,j_2)}_{\pm} 
(\mathbf{p},j,\l)]_{\a\b}\nonumber 
&=&\overline{{\mathcal U}}^\m_{\pm}\left(\mathbf{p},j,\l\right) 
[\overline{f}^{(j_1,j_2)} 
(p)]^{\a\b}{}_{\m}[f^{(j_1,j_2)}(p)]_{\a\b\n}{\mathcal U}_\nu \, _{\pm} 
(\mathbf{p},j,\l)\\ 
&=&\overline{{\mathcal U}}^\m_{\pm}(\mathbf{p},j,\l){\mathcal U}_\m\, _{\pm} 
(\mathbf{p},j,\l)=\pm 
1, 
\cec 
where $[\overline{w}^{(j_1,j_2)}_{\pm}(\mathbf{p},j,\l)]^{\a\b}=\g^0 
([w^{(j_1,j_2)}_{\pm}(\mathbf{p},j,\l)]^{\a\b})^\dagger\g^0$. 

Our pure spin-$\frac{3}{2}$ spinors, 
$\left[w^{\left( \frac{3}{2},0\right)}\left( {\mathbf p}, \frac{3}{2},\lambda \right)\right]^{\a\b}$, following from the equation 
(\ref{eq:atsvs}),  satisfy the condition 
\begin{equation} 
\gamma_\a\gamma_\b\left[w^{\left( \frac{3}{2},0\right)}\left( {\mathbf p}, \frac{3}{2},\lambda \right)\right]^{\a\b}=0, 
\end{equation} 
but are not solutions to the Dirac equation due to the non-commutativity of the $f$-tensors in (\ref{f3})--(\ref{eq:fintern320}) 
with    $p\cdot\gamma$. 

%%%%%%%%%%%%%%%%%%%%%%%%%%%%%%%%%%%%%%%%%%%%%%%%%%%%%%%%%%%%%%%%%%%%%%%%%%%%%% 

%%%%%%%%%%%%%%%%%%%%%%%%%%%%%%%%%%%%%%%%% 
%%%%%%%%%%%%%%%%%%%%%%%%%%%%%%%%%%%%%%%%%%%%%%%%%%%%%%%%%%%%%%%%%%%%%%%%%%%%%% 

%%%%%%%%%%%%%%%%%%%%%%%%%%%%%%%%%%%%%%%%% 
\section{\label{sec:5}Electromagnetic properties of  particles transforming according to 
the irreducible sectors of 
  the antisymmetric tensor-spinor space} 
%%%%%%%%%%%%%%%%%%%%%%%%%%%%%%%%%%%%%%%%%%%%%%%%%%%%%%%%%%%%%%%%%%%%%%%%%%%%%% 

%%%%%%%%%%%%%%%%%%%%%%%%%%%%%%%%%%%%%%%%% 
%%%%%%%%%%%%%%%%%%%%%%%%%%%%%%%%%%%%%%%%%%%%%%%%%%%%%%%%%%%%%%%%%%%%%%%%%%%%%% 

%%%%%%%%%%%%%%%%%%%%%%%%%%%%%%%%%%%%%%%%% 

In order to transfer the formalism to position space, 
one introduces plane waves of the type, 
$[\y^{(j_1,j_2);j}_\pm(x)]^{\a\b}=[w^{(j_1,j_2)}_\pm({\mathbf 
p},j,\l)]^{\a\b}e^{-ip\cdot x}$. 
In so doing, the momentum-space wave equations in (\ref{Gl_1})--(\ref{Gl_3}) 
amount to, 
\aeq\label{eq:eomps} 
\([\G^{(j_1,j_2);j}_{\m\n}]_{\a\b\g\d}\pd^\m \pd^\n+m^2 
\mathbf{1}_{\a\b\g\d}\) 
[\y^{(j_1,j_2);j}_\pm(x)]^{\g\d}=0, 
\ceq 
Minimal gauging is then standard and introduced by replacing ordinary by 
covariant derivatives, 
\aeq 
\pd^\m\longrightarrow D^\m= \pd^\m+ieA^\m 
\ceq 
where $e$ is the electric charge of the particle. 
%%%%%%%%%%%%%%%%%%%%%%%%%%%%%%%%%%%%%%%%%%%%%%%%%%%%%%%%%%%%%%%%%%%%%%%%%%%%%% 

%%%%%%%%%%%%%%%%%%%%%%%%%%%%%%%%%%%%%%%%% 

\subsection{\label{sec:5.1}The particle's Lagrangians} 
%%%%%%%%%%%%%%%%%%%%%%%%%%%%%%%%%%%%%%%%%%%%%%%%%%%%%%%%%%%%%%%%%%%%%%%%%%%%%% 

\begin{figure}
\centering{ 
\includegraphics{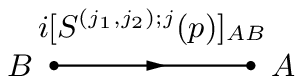} 
\caption{\label{propfig} Feynman rule for the propagators \eqref{eq:props} of 
particles in the 
$(j_2,j_1)\oplus(j_1,j_2)$ sector of spin-$j$.} 

\includegraphics{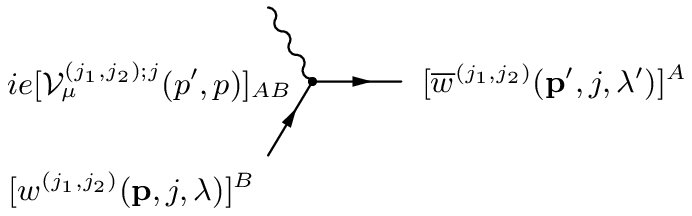} 
\caption{\label{regla1fig} Feynman rule for the one-photon vertex with 
$(j_2,j_1)\oplus(j_1,j_2)$ spin-$j$ particles in \eqref{vert1}.} 

\includegraphics{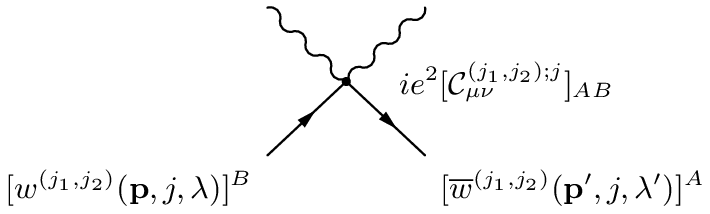} 
\caption{\label{regla2fig} Feynman diagram for the two-photon contact vertex 
with 
$(j_2,j_1)\oplus(j_1,j_2)$ spin-$j$ particles in \eqref{vert2}.}
} 
\end{figure}

%%%%%%%%%%%%%%%%%%%%%%%%%%%%%%%%%%%%%%%%% 
The free equations of motion \eqref{eq:eomps} for positive parity states can 
be now derived from the following Lagrangians: 
\aeq 
{\mathcal L}^{(j_1,j_2);j}_\textup{free}=(\pd^\m 
[\overline{\y}^{(j_1,j_2);j}]^A) 
[\G_{\m\n}^{(j_1,j_2);j}]_{AB}\pd^\n[\y^{(j_1,j_2);j}]^B 
-m^2[\overline{\y}^{(j_1,j_2);j}]^A[\y^{j_1,j_2);j}]_A, 
\quad A=\left[ \mu\nu\right],\quad B=\left[ \g\d\right], 
\ceq 
where we suppressed the arguments for the sake of simplifying notations. 
The gauged Lagrangians are then 
\aeq 
{\mathcal L}^{(j_1,j_2);j}=\left(D^{\m*} [\overline{\y}^{(j_1,j_2);j}]^A\right) 
[\G_{\m\n}^{(j_1,j_2);j}]_{AB}D^\n[\y^{(j_1,j_2);j}]^B 
-m^2[\overline{\y}^{(j_1,j_2);j)}]^A[\y^{j_1,j_2);j}]_A, 
\ceq 
whose decomposition into free and interacting parts is standard and reads, 
\aec 
{\mathcal L}^{(j_1,j_2);j}&=&{\mathcal L}^{(j_1,j_2);j}_\textup{free}+ 
{\mathcal L}^{(j_1,j_2);j}_\textup{int},\\ 
{\mathcal L}^{(j_1,j_2);j}_\textup{int}&=&- 
j_\m^{(j_1,j_2);j}A^\m+k_{\m\n}^{(j_1,j_2);j}A^\m A^\n. 
\cec 
Back to momentum space, we find 
\aec 
j_\m^{(j_1,j_2);j}({\mathbf p},\l, {\mathbf p}^\prime, 
\l^\prime)&=&e[\overline{w}_\pm^{(j_1,j_2)}(\mathbf{p}',j ,\l')]^A 
[\mathcal{V}^{(j_1,j_2);j}_\m(p',p)]_{AB} [w_\pm^{(j_1,j_2)} 
(\mathbf{p},j,\l)]^B,\\ 
k_{\m\n}^{(j_1,j_2);j}({\mathbf p},\l, {\mathbf p}^\prime, 
\l^\prime)&=&e^2[\overline{w}_\pm ^{(j_1,j_2)}(\mathbf{p}',j,\l')]^A 
[\mathcal{C}^{(j_1,j_2);j}_{\m\n}]_{AB} 
[w_\pm^{(j_1,j_2)}(\mathbf{p},j,\l)]^B, 
\cec 
the vertexes being given as, 
\aec 
~[\mathcal{V}^{(j_1,j_2);j}_\m(p',p)]_{AB}&=& 
[\G^{(j_1,j_2);j}_{\n\m}]_{AB}p'^\n+ 
[\G^{(j_1,j_2);j}_{\m\n}]_{AB}p^\n,\label{vert1}\\ 
~[\mathcal{C}^{(j_1,j_2);j}_{\m\n}]_{AB}&=&\frac{1}{2}\ 
([\G^{(j_1,j_2);j}_{\m\n}]_{AB}+[\G^{(j_1,j_2);j}_{\n\m}]_{AB} )\label{vert2}. 
\cec 
The Feynman rules following  from these Lagrangians are depicted in Figs. 
\ref{propfig}, \ref{regla1fig}, \ref{regla2fig}. 
%---------------------------Figures--------------------------------- 
%^^^^^^^^^^^^^^^^^^^^^^^^^^^^^^^^^^^^^^^^^^^^^^^^^^^^^^^^^^^^^^^^^^
It is not difficult to verify that the one-photon vertexes obey 
\aeq 
(p'-p)^\m [{\mathcal V}_\m^{(j_1,j_2);j}(p',p)]_{AB}=[B^{(j_1,j_2);j} 
(p')]_{AB}-[B^{(j_1,j_2);j}(p)]_{AB}. 
\label{WT_Id} 
\ceq 
which is the well known Ward-Takahashi identity, it is another sign of 
consistency of the interaction Lagrangians presented above, in particular, 
in can be used to proof gauge invariance of Compton scattering amplitudes 
\cite{DelgadoAcosta:2009ic}. 
%%%%%%%%%%%%%%%%%%%%%%%%%%%%%%%%%%%%%%%%%%%%%%%%%%%%%%%%%%%%%%%%%%%%%%%%%%%%%% 
%%%%%%%%%%%%%%%%%%%%%%%%%%%%%%%%%%%%%%%%% 
\subsection{\label{sec:5.2}Electromagnetic multipole moments } 
%%%%%%%%%%%%%%%%%%%%%%%%%%%%%%%%%%%%%%%%%%%%%%%%%%%%%%%%%%%%%%%%%%%%%%%%%%%%%% 

%%%%%%%%%%%%%%%%%%%%%%%%%%%%%%%%%%%%%%%%% 
%----------------------------------------------------------------------------- 

\subsubsection{\label{sec:5.2.1}Spin \texorpdfstring{$\frac{1}{2}$}{} in 
\texorpdfstring{$\left(\frac{1}{2},0\right)\oplus\left(0,\frac{1}{2}\right)$}{}} 
%----------------------------------------------------------------------------- 

The electromagnetic current, $ j_\mu^{(j_1,j_2);j}({\mathbf 
p}^\prime,\l^\prime, {\mathbf p},\l)$, 
of a particle with spin-$j$ residing in the $(j_1,j_2)\oplus (j_2,j_1)$ 
irreducible sector of the 
antisymmetric tensor-spinor space and between  states on 
their mass-shells for the case under consideration can be simplified  with the 
aid of the $f$-tensors in 
(\ref{f_tensors}) and cast exclusively  in terms of conventional Dirac-spinors 
as 
\aeq 
j_\m^{\left(\frac{1}{2},0\right);\frac{1}{2}}(\mathbf{p}',\l^\prime, 
\mathbf{p},\l) 
=e \overline{u}(\mathbf{p}',\l')(2m\g_\m)u(\mathbf{p},\l). 
\ceq 
This current is identical to Dirac's electromagnetic current and  implies 
precisely  same electromagnetic multipole moments as a 
genuine Dirac particle. 
%----------------------------------------------------------------------------- 

\subsubsection{\label{sec:5.2.2}Spin \texorpdfstring{$\frac{1}{2}$}{} in \texorpdfstring{$\left(\frac{1} 
{2},1\right)\oplus\left(1,\frac{1}{2}\right)$}{}} 
%----------------------------------------------------------------------------- 

In a way similar,  the current of the lower spin-$\frac{1}{2}$  companion to 
spin-$\frac{3}{2}$ in 
$\left( 
\frac{1}{2},1\right)\oplus\left(1,\frac{1}{2}\right)$ can be  calculated and 
simplified to express in terms of 
Dirac-spinors only with the result, 
\aeq 
j_\m^{\left(\frac{1}{2},1\right);\frac{1}{2}} 
(\mathbf{p}',\l',\mathbf{p},\l)=\frac{1}{3}e\overline{u} 
(\mathbf{p}',\l')(4(p'+p)^\m-2m \g_\m)u(\mathbf{p},\l). 
\ceq 
As expected, also this bilinear is nothing else but the electromagnetic current of spin- 
$\frac{1}{2}$ in the non-Dirac sector 
of the four-vector spinor. 
%, as  earlier reported in \cite{Delgado-Acosta:2013kra}. 
%----------------------------------------------------------------------------- 

\subsubsection{\label{5.2.3}Spin \texorpdfstring{$\frac{3}{2}$}{} in 
\texorpdfstring{$\left (\frac{1}{2},1\right)\oplus\left(1,\frac{1}{2}\right)$}{}}
%----------------------------------------------------------------------------- 

{}For this case, the electromagnetic current is calculated to take the 
following form, 
\aeq 
j_\m^{\left(\frac{1}{2},1\right);\frac{3}{2}} 
(\mathbf{p}',\l^\prime,\mathbf{p},\l )=e\overline{{\mathcal U}}^\a_+ 
\left(\mathbf{p}',\frac{3}{2},\l'\right)(2m g_{\a\b}\g_\m){{\mathcal U}}^\b 
_+\left(\mathbf{p},\frac{3}{2},\l\right), 
\ceq 
where ${\mathcal U}^\b_+\left( \mathbf{p},\frac{3}{2},\l\right)$ are the 
conventional positive 
parity states in the Rarita-Schwinger four vector-spinor, meaning that the 
electromagnetic  moments under discussion are identical to those of a Rarita- 
Schwinger particle (see Eq. (4.28) in \cite{DelgadoAcosta:2012yc}) and read,
\begin{subequations} 
\label{eq:emmrs} 
\aec 
~[Q_E^0(\l)]_{RS}&=&e,\\ 
~[Q_M^1(\l)]_{RS}&=&\frac{2}{3}\(\frac{e}{2m}\)\langle 
S_z\rangle,\label{magdiprs}\\ 
~[Q_E^2(\l)]_{RS}&=&\frac{1}{3}\(\frac{e}{m^2}\)\langle {\mathcal A}\rangle,\\ 
~[Q_M^3(\l)]_{RS}&=&2\(\frac{e}{2m^3}\)\langle {\mathcal B}\rangle . 
\cec 
\end{subequations} 
Here, $Q_E^2(\l)$ and $Q_M^3(\l) $ in turn denote the electric quadrupole, 
and magnetic octupole 
moments. Their explicit values correspond to a particular 
polarization are found from $\langle {\mathcal O}\rangle\equiv \langle \l 
\vert 
{\mathcal O}\vert \l\rangle$ by inserting the explicit form of the ${\mathcal 
A}$ and ${\mathcal B}$ operators (in obvious notations): 
\aec 
{\mathcal A}&=&3S_z^2-\mathbf{S}^2,\\ 
{\mathcal B}&=&S_z\(15 S_z^2-\frac{41}{5}\mathbf{S}^2\). 
\cec 
The gyromagnetic factor associated with \eqref{eq:emmrs} can only be 
identified 
via its dipole magnetic moment, $Q_M^1(\lambda)$,  in \eqref{magdiprs} as 
$g_{RS}=\frac{2}{3}$. 
Therefore, in combination with the result of the preceding sub-subsection , 
our method correctly describes the Rarita-Schwinger four-vector spinor 
sector of the antisymmetric tensor-spinor space. However within the more 
general method of the 
Poincar\'e covariant projectors for spin-$\frac{3}{2}$ description within the 
four-vector spinor 
space \cite{Napsuciale:2006wr}, the gyromagnetic ratio, $g$,  is identified at 
the level of the 
current and all the electromagnetic moments depend only on this parameter, 
because the 
currents within this method exhibit the general structure of  two-term Gordon- 
decompositions. 
The general expressions reported within the latter method are essentially 
different 
from \eqref{eq:emmrs}, specially in the relations between the highest moments 
with respect to the dipole magnetic moment, they can be found in 
\cite{DelgadoAcosta:2012yc} and read 
\begin{subequations} 
\label{eq:emmvs} 
\aec 
~[Q_E^0(\l)]_{VS}&=&e,\\ 
~[Q_M^1(\l)]_{VS}&=&g\(\frac{e}{2m}\)\langle S_z\rangle,\\ 
~[Q_E^2(\l)]_{VS}&=&(1-g)\(\frac{e}{m^2}\)\frac{1}{3}\langle {\mathcal 
A}\rangle,\\ 
~[Q_M^3(\l)]_{VS}&=&g\(\frac{e}{2m^3}\)\langle {\mathcal B}\rangle. 
\cec 
\end{subequations} 
The $g$ value for the highest spin $\frac{3}{2}$ within the four-vector spinor space 
has been fixed in \cite{Napsuciale:2006wr} to  $g=2$ 
from the requirement on causality of propagation within an electromagnetic environment. 

%----------------------------------------------------------------------------- 

\subsubsection{\label{sec:5.2.4}Spin \texorpdfstring{$\frac{3}{2} $}{} in \texorpdfstring{$\left (\frac{3} 
{2},0\right)\oplus\left(0,\frac{3}{2}\right)$}{}} 
%----------------------------------------------------------------------------- 

Executing the same strategy as in the previous subsections, we calculate 
the electromagnetic  multipole moments of  a particle transforming in the 
single spin $\frac{3}{2}$ irreducible Weinberg-Joos sector of the 
antisymmetric tensor spinor of interest. In this case the current reads 
\aeq\label{current-s32} 
j_{\mu}^{\left(\frac{3}{2},0\right);\frac{3}{2}} 
(\mathbf{p}',\l',\mathbf{p},\l)= 
e\left[\overline{w}^{(\frac{3}{2},0)}(\mathbf{p}',3/2,\l')_+\right]^A 
\left[ \mathcal{V}_\m^{\left(\frac{3}{2},0\right);\frac{3}{2}} 
(p',p)\right]_{AB} 
\left[{w}^{(\frac{3}{2},0)}(\mathbf{p},3/2,\l)_+\right]^B 
\ceq 
with the $\mathcal{V}_\m^{\left(\frac{3}{2},0\right);\frac{3}{2}}(p',p)$ vertex given in \eqref{vert1}. The latter has been expressed in terms of the 
$[\G^{\left(3/2,0\right);3/2}_{\m\n}]_{AB}$ 
tensor in \eqref{gamma-s32},  and of the Lorentz-invariant irrep projector, $[\mathcal{P}^{(3/2,0)}_F]_{AB}$, 
in \eqref{CS_3}. 
This current  can be further simplified in taking advantage of 
the equation \eqref{eq:atsvs}, 
which relates tensor-spinors to vector-spinors. In so doing,  the current 
in \eqref{current-s32}  re-expresses in terms of vector-spinors as: 
\begin{eqnarray} 
j_{\mu}^{\left(\frac{3}{2},0\right);\frac{3}{2}} 
(\mathbf{p}',\l',\mathbf{p},\l)&=& 
\frac{38}{9}\,e\, 
\overline{{\mathcal U}}^\a_+ 
\left(\mathbf{p}',\frac{3}{2},\l'\right)\left( 
(p'+p)_\m g_{\a\b}-m g_{\a\b}\g_\m-(p'_\b g_{\a\m}+p_\a 
g_{\b\m})\right.\nonumber\\ 
&&\left.+\frac{20}{38 m}( p_\a p'_\b-p'\cdot p \,g_{\a\b})\g_\m \right) 
{\mathcal U}^\b _+\left(\mathbf{p},\frac{3}{2},\l\right). 
\end{eqnarray} 
The procedure to find the multipole moments from known currents is 
well established  (see for example \cite{Lorce:2009bs},\cite{DelgadoAcosta:2012yc} and references therein) and amounts to 
\begin{subequations} 
\label{eq:emmss32} 
\aec 
~[Q_E^0(\l)]&=&e,\\ 
~[Q_M^1(\l)]&=&\frac{2}{3}\(\frac{e}{2m}\)\langle S_z\rangle,\\ 
~[Q_E^2(\l)]&=&-\frac{1}{3}\(\frac{e}{m^2}\)\langle {\mathcal A}\rangle,\\ 
~[Q_M^3(\l)]&=&-2\(\frac{e}{2m^3}\)\langle {\mathcal B}\rangle. 
\cec 
\end{subequations} 
The latter expressions  fully coincide in form  with those  earlier reported in 
\cite{DelgadoAcosta:2012yc} and equivalent to the Weinberg-Joos formalism 
where the calculation have been carried out while treating the states under 
consideration as eight-component vectors. 
The difference is that here the gyromagnetic ratio is 
fixed to the inverse of the spin, 
$g=\frac{2}{3}$, and in accord with Belinfante's conjecture, while in 
\cite{DelgadoAcosta:2012yc},
%\footnote{Please see only Table 1 of \cite{DelgadoAcosta:2012yc} and this time ignore Eq. (4.52) in \cite{DelgadoAcosta:2012yc}, which has two %typos; the $g$-factor symbol in the expressions for the magnetic dipole and octupole moments is missing, it should be there.}
where only a covariant spin-projector has been used, 
$g$ had remained unspecified according to (see Table 1 in \cite{DelgadoAcosta:2012yc}), 
\begin{subequations} 
\label{eq:emmts} 
\aec 
~[Q_E^0(\l)]_{TS}&=&e,\\ 
~[Q_M^1(\l)]_{TS}&=&g\(\frac{e}{2m}\)\langle S_z\rangle,\\ 
~[Q_E^2(\l)]_{TS}&=&-(1-g)\(\frac{e}{m^2}\)\langle {\mathcal A}\rangle,\\ 
~[Q_M^3(\l)]_{TS}&=&-3 g\(\frac{e}{2m^3}\)\langle {\mathcal B}\rangle. 
\cec 
\end{subequations} 
We conclude that the antisymmetric tensor-spinor is perfectly  well suited for 
the  adequate  description of particles of 
spin-$\frac{3}{2}$ transforming in the  single-spin valued Weinberg-Joos 
representation space. 
Notice difference between the sets of observables in (\ref{eq:emmts}) and 
(\ref{eq:emmvs}). 
%%%%%%%%%%%%%%%%%%%%%%%%%%%%%%%%%%%%%%%%%%%%%%%%%%%%%%%%%%%%%%%%%%%%%%%%%%%%%% 

%%%%%%%%%%%%%%%%%%%%%%%%%%%%%%%%%%%%%%%%% 
\section{\label{sec:6}Compton scattering off spin-\texorpdfstring{$\frac{3}{2}$}{} particles in 
\texorpdfstring{$\left(\frac{3}{2},0\right)\oplus\left(0,\frac{3}{2}\right)$}{}} 
%%%%%%%%%%%%%%%%%%%%%%%%%%%%%%%%%%%%%%%%%%%%%%%%%%%%%%%%%%%%%%%%%%%%%%%%%%%%%% 
%%%%%%%%%%%%%%%%%%%%%%%%%%%%%%%%%%%%%%%%% 
The formalism developed in the present work provides a well defined and comfortably managable 
technical tool for  calculations of scattering cross sections in terms  of matter fields as Lorentz tensors, 
thus avoiding the cumbersome and computer time consuming  matrix spinor calulus, which for the specific case under 
consideration would require the construction of $8\times 8$ matrix invariants for diagonal processes, or, rectangular 
$4\times 8$ bilinears for spin-$\frac{1}{2}\to \frac{3}{2}$ transitions. 
Above, we employed  this tool to calculate  the electromagnetic 
multipole moments of  all the particles populating the irreducible sectors of the antisymmetric-tensor spinor space. 
However, the latter properties characterize the particles when they are at rest, while one also wants to know 
how they behave in dynamical processes such as collisions. {}For this purpose, we apply the method suggested 
in the  study of processes  involving particles in flight, as is the Compton scattering, the subject of this section. 

\begin{figure}
\centering{ 
\includegraphics{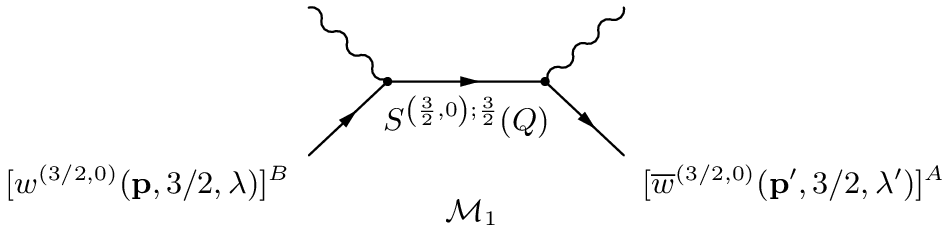} 
\caption{\label{mafig} Diagram for the direct-scattering contribution 
\eqref{maj1} to the Compton scattering amplitude \eqref{camp}.} 

\includegraphics{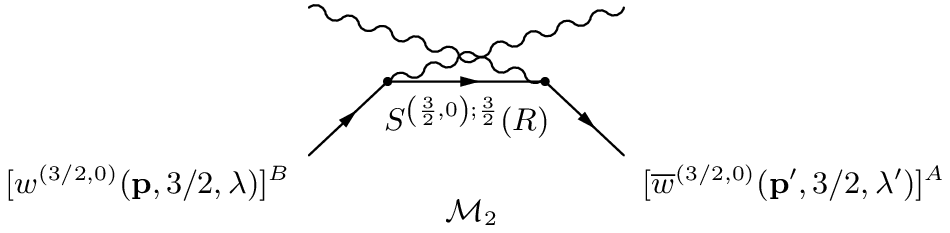} 
\caption{\label{mbfig} Diagram for the exchange-scattering contribution 
\eqref{mbj1} to the Compton scattering amplitude \eqref{camp}.} 

\includegraphics{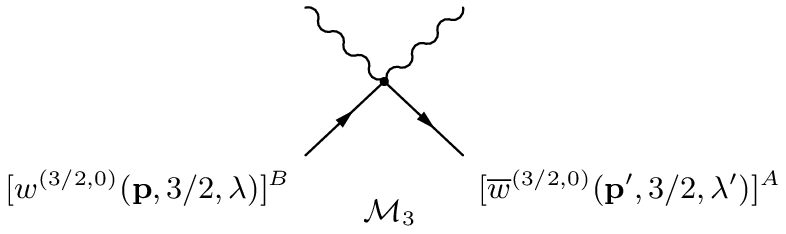} 
\caption{\label{mcfig} Diagram for the point-scattering contribution 
\eqref{mcj1} to the Compton scattering amplitude \eqref{camp}.}
} 
\end{figure}

\begin{figure}[ht]
\centering{ 
\includegraphics{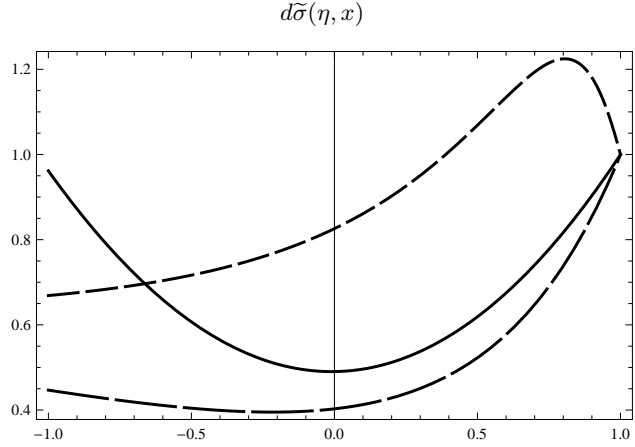} 
\caption{\label{dsfig} The differential cross section, $d\widetilde{\s}(\h,x)$, as a 
function of $x=\cos\q$. The solid curve represents the classical limit at 
$\h=\o/m=0$, the long dashed line corresponds to an energy comparable to the mass 
of the particle, $\h= 1$, while the short dashed curve corresponds to $\h=2.5$. 
This cross section increases with energy, except in the forward 
direction, $x=1$, where it approaches  $d\widetilde{\s}(\h,1)=1$.}
} 
\end{figure}

The tree-level Compton scattering amplitude  \cite{Bjorken} contains contributions from 
three different channels (see Figs. \ref{mafig}, \ref{mbfig}, \ref{mcfig}). 
%---------------------------Figures-------------------- 
%---------------------------Figures-- 
Here, $p$ and $p'$ to denote in turn the four-momenta of the incident and scattered 
single spin-$\frac{3}{2}$ 
target particles, while  $q$ and $q'$ are  the four-momenta of the incident 
and scattered photons, respectively. Then the amplitude has the following form 
\aeq\label{camp} 
\mathcal{M}=\mathcal{M}_1+\mathcal{M}_2+\mathcal{M}_3, 
\ceq 
where 
\aec 
i\mathcal{M}_1&=&e^2\left[\overline{w}^{\left(\frac{3} 
{2},0\right)}\left(\mathbf{p}',\frac{3}{2},\l'\right)\right]^{A}~ 
\left[\mathcal{U}_{\m\n}(p',Q,p)\right]_{AB} 
\left[ 
w^{\left(\frac{3}{2},0\right)} 
\left(\mathbf{p},\frac{3}{2},\l\right)\right]^{B}[\e^\m(\mathbf{q}',\ell')]^* 
\e^\n(\mathbf{q},\ell),\label{maj1}\\ 
i\mathcal{M}_2&=&e^2\left[\overline{w}^{\left(\frac{3}{2},0\right)} 
\left(\mathbf{p}',\frac{3}{2},\l'\right)\right]^{A}~ 
[\mathcal{U}_{\n\m}(p',R,p)]_{AB} 
\left[ w^{\left(\frac{3}{2},0\right)}\left(\mathbf{p},\frac{3} 
{2},\l\right)\right]^{B}[\e^\m(\mathbf{q}'.\ell')]^* 
\e^\n(\mathbf{q},\ell),\label{mbj1}\\ 
-i\mathcal{M}_3&=&e^2\left[ 
\overline{w}^{\left(\frac{3}{2},0\right)}\left(\mathbf{p}',\frac{3} 
{2},\l'\right)\right]^{A} 
~[\mathcal{X}_{\m\n}]_{AB}[\e^\m(\mathbf{q}',\ell')]^* 
\e^\n(\mathbf{q},\ell)\label{mcj1}, 
\cec 
with $Q=p+p'=q+q'$ and $R=p'-q=p-q'$, and 
\aec 
~[\mathcal{U}_{\m\n}(p',Q,p)]_{AB}&=&\left[\mathcal{V}^{\left(\frac{3} 
{2},0\right);\frac{3}{2}}_{\m} 
(p',Q)\right]_{AC} 
~\left[S^{\left(\frac{3}{2},0\right);\frac{3}{2}} 
(Q)\right]^{CD}\left[\mathcal{V}^{\left(\frac{3}{2},0\right);\frac{3} 
{2}}_\n(Q,p)\right]_{DB},\\ 
~[\mathcal{X}_{\m\n}]_{AB}&=&\left[\mathcal{C}_{\m\n}^{\left(\frac{3} 
{2},0\right);\frac{3}{2}}+\mathcal{C}_{\n\m}^{\left( \frac{3} 
{2},0\right);\frac{3}{2}}\right]_{AB} 
\cec 
The gauge invariance of this amplitude is ensured by the Ward-Takahashi 
identity 
(see for example \cite{DelgadoAcosta:2009ic}). The averaged squared amplitude 
is then: 
\aec\label{m2av} 
\overline{\left\vert\mathcal{M}\right\vert^2}&=&\frac{1} 
{4}\sum_{\l,\l',\ell,\ell'} 
\mathcal{M}[\mathcal{M}]^\dagger\\ 
&=&Tr\left[[\mathcal{M}_{\m\n}(p',Q,R,p)]_{AB} 
[\mathcal{M}^{\n\m}(p,R,Q,p')]^{AB}\right] \label{eq:asa} 
\cec 
where we have defined 
\aec 
~[\mathcal{M}_{\m\n}(p',Q,R,p)]_{AB}&=& 
\frac{e^2}{2}[\mathbb{P}^{\left(\frac{3}{2},0\right);\frac{3}{2}}_+ 
(\mathbf{p}')]_{A} 
{}^C[\mathbb{U}_{\m\n}(p',Q,R,p)]_{CB},\\ 
~[\mathbb{U}_{\m\n}(p',Q,R,p)]_{CB}&=&[\mathcal{U}_{\m\n} 
(p',Q,p)+\mathcal{U}_{\n\m}(p',R,p)-{\mathcal{X}}_{\m\n}]_{CB}. 
\cec 
We furthermore used 
\aec 
\sum_{\ell}\e^\m(\mathbf{q},\ell)[\e^\n(\mathbf{q},\ell)]^{*}&=&- 
g^{\m\n}\label{phpr},\\ 
\sum_{\l}\left[ w_+^{\left(\frac{3}{2},0\right)}\left(\mathbf{p},\frac{3} 
{2},\l\right)\right]_{A} 
\left[\overline{w}_+^{\left(\frac{3}{2},0\right)} 
\left(\mathbf{p},\frac{3}{2},\l\right)\right]_{B} 
&=&[\mathbb{P}^{\left(\frac{3}{2},0\right);\frac{3}{2}}_+(\mathbf{p})]_{AB}, 
\cec 
where the projector over spin-$\frac{3}{2}$ states with positive parity can be 
shown 
to have the following form 
\aeq 
~\left[\mathbb{P}_{+}^{\left(\frac{3}{2},0\right);\frac{3}{2}} 
(\mathbf{p})\right]_{AB} 
=\left[f^{\left(\frac{3}{2},0\right)}(\mathbf{p})\right]_{A}{}^{\m}\(\frac{- 
\slashed{p}+m}{2m}\) 
\left[\overline{f}^{\left(\frac{3}{2},0\right)}(\mathbf{p})\right]_{B\m}. 
\ceq 
The contractions indicated in \eqref{eq:asa} are easier performed with the aid 
of the FeynCalc package 
giving as a result the following expression: 
\begin{eqnarray}\label{m2nkr12g} 
\overline{\vert{\mathcal M}\vert^2}&=& 
\frac{1}{162 m^6 \left(m^2-s\right)^2 \left(m^2-u\right)^2} 
\sum_{k=1}^7 m^{2k} a_{2k}, 
\end{eqnarray} 
where $s,u$ are the standard Mandelstam variables and we are using the 
notations
\begin{subequations} 
\aec 
a_0&=&18 s^2 u^2 (s+u)^3,\\ 
a_2&=&-9 s u (s+u)^2 (7 (s^2+u^2)+8 s u),\\ 
a_4&=&(s+u) (63 (s^4+u^4)+348 (s^3 u+s u^3)+578 s^2 u^2),\\ 
a_6&=&-165 (s^4+u^4)-588 (s^3 u+s u^3)-574 s^2 u^2,\\ 
a_8&=&2 (s+u) (5 (s^2+u^2)-142 s u),\\ 
a_{10}&=&2 (105 (s^2+u^2)-158 s u),\\ 
a_{12}&=&-280 (s+u),\\ 
a_{14}&=&912. 
\cec
\end{subequations} 
Now we can obtain the differential cross section in the laboratory frame from 
the standard formulas 
\aec 
\frac{d\s}{d\Omega}&=&\(\frac{1}{8\p m}\frac{\o'} 
{\o}\)^2\overline{\vert{\mathcal M}\vert^2},\\ 
\o'&=&\frac{m\o}{m+(1-\cos\q)\o}, 
\cec 
where $\o$ and $\o'$ are the energies of the incident and scattered photons 
respectively, while  $\q$ is 
the scattering angle in the laboratory frame. Furthermore, with 
\aec 
s&=&m(m+2\o),\\ 
u&=&m(m-2\o'), 
\cec 
and after some algebraic manipulations,  the final result can be given the form of an expansion in powers of $\eta=\o/m$ according to, 
\aeq\label{dss} 
\frac{d \s(\h,x)}{d \Omega}=\frac{r_0^2}{162 (\eta  (x-1)-1)^5} 
\sum_{k=0}^6 \h^k b_k, 
\ceq 
with $r_0=e^2/(4\p m)=\a m$, $x=\cos\q$, and the expansion coefficients being, 
\begin{align} 
b_0=&-81 (x^2+1),\\ 
b_1=&243 (x-1) (x^2+1),\\ 
b_2=&-(x-1) (243 x^3-333 x^2+338 x-468),\\ 
b_3=&(x-1)^2 (81 x^3-261 x^2+271 x-531),\\ 
b_4=&(x-1)^2 (90 x^3-233 x^2+440 x-459),\\ 
b_5=&6 (x-1)^3 (8 x^2-20 x+39),\\ 
b_6=&9 (x-1)^3 (x^2-5 x+8). 
\end{align} 
In the low energy limit, we recover as expected the Thompson differential 
cross 
section: 
\aeq 
\lim_{\h\rightarrow 0}\frac{d \s(\h,x)}{d \Omega}=\frac{1}{2} r_0^2 
\left(x^2+1\right), 
\ceq 
while in forward direction, the differential cross section takes an energy 
independent value, 
\aeq 
\lim_{x\rightarrow 1} \frac{d \s(\h,x)}{d \Omega}=r_0^2, 
\ceq 
and in accord with unitarity. 
In all other directions however, the differential cross section increases with 
energy. In the Figure \ref{dsfig} we present a plot of the quantity
\aeq
d\widetilde{\s}(\h,x)\equiv \frac{1}{r_0^2}\frac{d \s(\h,x)}{d \Omega},
\ceq 
as a function of the $x=\cos\q$ variable, at energies of $\h=0$ (solid curve), $\h=1$ (long dashed curve) and $\h=2.5$ (short dashed curve), here we see how the differential cross section approaches the classical limit at low energy (symmetric curve) and raises as the energy grows except in the forward direction.

\begin{figure}[ht]
\centering{ 
\includegraphics{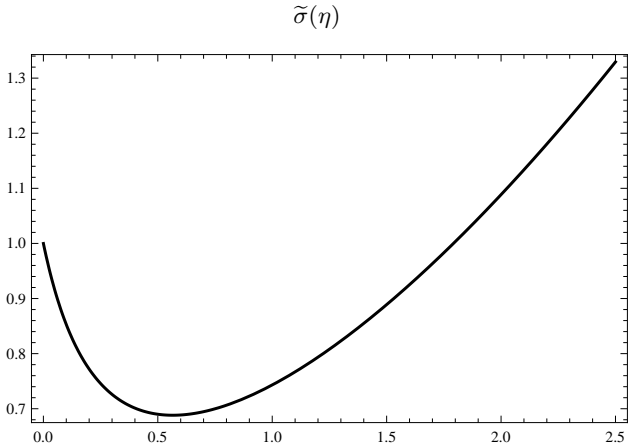} 
\caption{\label{sfig} The total cross section $\widetilde{\s}(\h)$ as a function 
of $\h=\o/m$. In the low energy limit the Thompson limit, 
$\widetilde{\s}(0)=1$, is recovered, otherwise the cross section 
grows with the energy increase.}
} 
\end{figure}

Integrating over the solid angle we find the total cross section as: 
\aeq\label{tcs} 
\s(\h)=\sum_{k=0}^8 \frac{\h^k c_k\s_T}{108 \eta ^2 (2 \eta +1)^4} 
+\sum_{\ell=0}^4 \frac{\h^\ell h_\ell\s_T \log (2 \eta +1)}{216 \eta ^3}, 
\ceq 
being $\s_T=(8/3)\pi r_0^2$ the Thompson cross section and 
%\aec 
%c_0&=&162,\\ 
%c_1&=&1566,\\ 
%c_2&=&6217,\\ 
%c_3&=&12796,\\ 
%c_4&=&14244,\\ 
%c_5&=&8011,\\ 
%c_6&=&1794,\\ 
%c_7&=&126,\\ 
%c_8&=&72,\\ 
%h_0&=&-162,\\ 
%h_1&=&-432,\\ 
%h_2&=&-277,\\ 
%h_3&=&-21,\\ 
%h_4&=&27. 
%\cec
\begin{subequations} 
\begin{align} 
c_0&=162, &c_1=&1566,\\ 
c_2&=6217,&c_3=&12796,\\ 
c_4&=14244,&c_5=&8011,\\ 
c_6&=1794,&c_7=&126,\\ 
c_8&=72,&h_0=&-162,\\ 
h_1&=-432,&h_2=&-277,\\ 
h_3&=-21,&h_4=&27. 
\end{align} 
\end{subequations}
The total cross section \eqref{tcs} has the following limits, 
\aec 
\lim_{\h\rightarrow 0}\s(\h)&=&\s_T,\\ 
\lim_{\h\rightarrow 
\infty}\s(\h)&=&\infty. 
\cec 
This behavior is shown in the Figure \ref{sfig}, where we make a plot of
\aeq
\widetilde{\s}(\h)\equiv \frac{\s(\h)}{\s_T},
\ceq
here we see the decreasing behavior of the cross section at low energies as well as its growing behavior at high energies.

\section{\label{sec:7}Conclusions} 
In the present work we constructed the physical equivalent to the Weinberg- 
Joos theory for single spin-$j$ particles by replacing the 
multi-component spinors by Lorentz tensors for bosons, or tensor-spinors for 
fermions and the higher order differential wave equations by 
such of second order. The theory is based on the relativistic invariants (RInS) 
of both the Lorentz- and Poincar\'e algebras, as well as on the fact that any irreducible 
sector of any Lorentz tensor, with or without Dirac spinor 
components, is equally good for the description of the elementary particle 
residing in it as a single irreducible representation. Stated 
differently, the individuality of the fundamental particles residing within 
given irreducible representation spaces of the Lorentz 
algebra are fully respected by any direct sum of them. 
Indeed,  our approach, which we illustrated 
for the sake of concreteness and without loss of generality 
on the example of spin-$\frac{3}{2}$ in 
$\left[ 
\left( 1,0\right)\oplus\left(0,1\right)\right] 
\oplus \left[\left(\frac{1}{2},0 \right)\oplus \left( 0,\frac{1}{2}\right)\right]$, 
correctly reproduces the electromagnetic multipole 
moments of the particles 
in each one of the irreducible sectors of the anti-symmetric tensor spinor 
space, be them the Dirac, 
the Rarita-Schwinger, or the pure spin-$\frac{3}{2}$ sectors, 
$\left(\frac{3}{2},0 \right)\oplus\left(0,\frac{3}{2} \right)$. 
We were able to show that unitarity 
is respected in  Compton scattering off pure spin-$\frac{3}{2}$ 
in forward direction within a minimal gauging scheme and without 
any need of invoking non-minimal couplings, and in parallel to the same behaviour of 
spin-$\frac{3}{2}$ transforming within the four-vector spinor \cite{DelgadoAcosta:2009ic}. However,  the gyromagnetic ratio 
for the case considered here has been found as the inverse of 
the spin, thereby matching Belinfante's conjecture rather than the 
universal $g=2$ value established for particles transforming as 
the highest spins in irreps of multiple spins and parities. 
This finding emphasizes once again the observation that fundamental 
particles residing in non-equivalent $so(1,3)$ representation spaces 
can be equipped by distinct physical properties and are likely to participate 
in different physical processes. The scheme elaborated here  allows 
detecting such differences, is friendly towards symbolic computational softwares such as Mathematica and FeynCalc, 
and significantly less computer time consuming  than the conventional matrix spinor calculus. 
%%%%%%%%%%%%%%%%%%%%%%%%%%%%%%%%%%%%%%%%%%%%%%%%%%%%%%%%%%%%%%%%%%%%%%%%%%%%%% 

%%%%%%%%%%%%%%%%%%%%%%%%%%%%%%%%%%%%%%%%% 

\end{document}